\documentclass{iopart}
\usepackage{iopams}

\usepackage[english]{babel}
\usepackage{latexsym}
\usepackage{graphics}
\usepackage{graphicx}
\usepackage{subfigure}
\usepackage{epsfig}

\newcommand{\epsfboxmod}[1]{\epsfbox{#1.eps}}

\newcommand{\infig}[2]{\begin{center}
                                    \mbox{ \epsfxsize #1 \epsfboxmod{#2}}
                                      \vspace{-0.8cm}
                                    \end{center}}

\begin{document}

\renewcommand{\i}{\textrm{i}}
\newcommand{\ie}{{\it i.e. }}
\newcommand{\eg}{{\it e.g. }}

\newcommand{\erfc}{\textrm{erfc}}
\newcommand{\sinc}{\textrm{sinc}}

\newcommand{\Vopt}{V}
\newcommand{\tVopt}{\widetilde{V}}
\newcommand{\Vr}{V_\textrm{\tiny R}}
\newcommand{\tVr}{\tilde{V}_\textrm{\tiny R}}
\newcommand{\sigmar}{\sigma_\textrm{\tiny R}}

\newcommand{\Npeaks}{N_\textrm{\tiny peaks}}

\newcommand{\xiini}{\xi_\textrm{\tiny in}}
\newcommand{\nc}{n_\textrm{\tiny c}}
\newcommand{\gammaeff}{\gamma_\textrm{\tiny eff}}
\newcommand{\Lloc}{L_\textrm{\tiny loc}}

\newcommand{\kc}{k_\textrm{\tiny c}}

\newcommand{\muTF}{\mu_\textrm{\tiny TF}}
\newcommand{\LTF}{L_\textrm{\tiny TF}}

\newcommand{\Ekin}{E_\textrm{\tiny kin}}
\newcommand{\Eint}{E_\textrm{\tiny int}}

\newcommand{\asc}{a_\textrm{\tiny sc}}
\newcommand{\goned}{g}
\newcommand{\ldB}{\lambda_\textrm{\tiny dB}}

\title[Disorder-induced trapping versus Anderson localization in expanding BECs]
{Disorder-induced trapping versus Anderson localization in Bose-Einstein condensates expanding in disordered potentials}

\author{L.~Sanchez-Palencia, D.~Cl\'ement, P.~Lugan, P. Bouyer, and A. Aspect}
\address{Laboratoire Charles Fabry de l'Institut d'Optique,
CNRS and Univ. Paris-Sud,
Campus Polytechnique, RD 128, 
F-91127 Palaiseau cedex, France}

\begin{abstract}
We theoretically investigate the localization of an expanding Bose-Einstein condensate with
repulsive atom-atom interactions in a disordered potential. 
We focus on the regime where the initial inter-atomic interactions dominate over the kinetic energy
and the disorder.
At equilibrium in a trapping potential and for the considered small disorder,
the condensate shows a Thomas-Fermi shape modified by the disorder.
When the condensate is released from the trap, a strong suppression of the expansion is obtained in contrast
to the situation in a periodic potential with similar characteristics.
This effect crucially depends on both the momentum distribution of the expanding BEC and
the strength of the disorder.
For strong disorder as in the experiments reported by
D.~Cl\'ement {\it et al.}, Phys. Rev. Lett. {\bf 95}, 170409 (2005)
and
C.~Fort {\it et al.}, Phys. Rev. Lett. {\bf 95}, 170410 (2005),
the suppression of the expansion results from the fragmentation of the core of the condensate and
from classical reflections from large modulations of the disordered potential in the tails of the condensate.
We identify the corresponding disorder-induced trapping scenario for which large atom-atom interactions and
strong reflections from single modulations of the disordered potential play central roles.
For weak disorder, the suppression of the expansion signals the onset of Anderson localization,
which is due to multiple scattering from the modulations of the disordered potential.
We compute analytically the localized density profile of the condensate and show that the localization
crucially depends on the correlation function of the disorder. In particular, for speckle potentials
the long-range correlations induce an effective mobility edge in 1D finite systems.
Numerical calculations performed in the mean-field approximation support our analysis
for both strong and weak disorder.
\end{abstract}

\date{\today}


\tableofcontents

%%%%%%%%%%%%%%%%%%%%%%%%%%%%%%%%%%%%%%%%%%%%%%%%%%%%%%%%%%%%%%%%%%%%%%
\section{Introduction}
\label{introduction}

	%%%%%%%%%%%%%%%%%%%%%%%%%%%%%%%%%%%
	\subsection{Disorder and ultracold atomic gases}
\label{introduction.disorder}

Understanding the effect of disorder in physical systems is of fundamental importance
in various domains, such as
mechanics,
wave physics,
solid-state physics,
quantum fluid physics, or
atomic physics.
Although in many situations this effect is weak and can be ignored in first approximation,
it is not always so.
Strikingly enough, even arbitrarily weak disorder can dramatically change the properties of physical
systems and result in a variety of
non-intuitive phenomena. Many of them are not yet fully understood.
Examples in classical systems include
Brownian motion \cite{risken1989},
percolation \cite{aharony1994},
and magnetism in dirty spin systems \cite{imry1975,imbrie1984,bricmont1987,aizenman1989,aizenman1990}.
In quantum systems the effects of disorder can be particularly strong owing to the
complicated interplay of interference, particle-particle interactions and disorder.
The paradigmatic example is (strong) Anderson localization of non-interacting particles \cite{anderson1958,nagaoka1982,ando1988,vantiggelen1999}.
Other interesting effects of disorder in quantum systems include
weak localization and coherent back-scattering \cite{akkermansbook},
disorder-driven quantum phase transitions and the corresponding
Bose glass \cite{giamarchi1988,fisher1989,scalettar1991} and
spin glass \cite{parisi1987,sachdev1999} phases.

Anderson localization (AL) signals out in two equivalent ways,
either as the suppression of the transport of matterwaves in disordered media,
or as an exponential decay at large distances of the envelope of the eigenstates of free-particles
in a disordered potential \cite{vantiggelen1999}.
Both properties strongly contrast with the case of periodic potentials,
in which transport is free and
all eigenstates extend over the full system as demonstrated by the Bloch theorem
\cite{ashcroft1976}.
Anderson localization is due to a destructive
interference of particles (waves) which multiply scatter
from the modulations of a disordered potential. It is thus
expected to occur when interference plays a central role in
the multiple scattering process \cite{vantiggelen1999}.
In three dimensions,
it requires the particle wavelength to be larger than the
scattering mean free path $l$ as pointed out by Ioffe and
Regel \cite{ioffe1960}. One then finds a mobility edge at momentum
$k=1/l$, below which AL can appear. In one and two
dimensions all single-particle quantum states are predicted
to be localized \cite{mott1961,thouless1977,gang4}, although for certain types of disorder
one has an effective mobility edge in the Born approximation \cite{izrailev1999,izrailev2005,lsp2007}.

Ultracold atomic gases are now widely considered to revisit standard problems
of condensed matter physics under unique control possibilities.
Dilute atomic Bose-Einstein condensates (BEC) \cite{cornell2001,ketterle2001,dalfovo1999,pitaevskii2004} and
degenerate Fermi gases (DFG) \cite{truscott2001,schreck2001,hadzibabic2002,roati2002,giorgini2007}
are produced routinely taking advantage of
the recent progress in cooling and trapping of neutral atoms \cite{chu1997,cct1997,phillips1997}.
In addition, controlled potentials with no defects, for instance periodic potentials (optical lattices),
can be designed in a large variety of geometries \cite{lattices2000}.
In periodic optical lattices, transport has been widely
investigated, showing lattice-induced reduction of mobility
\cite{burger2001,kramer2002,fertig2005} and
interaction-induced self-trapping \cite{trombettoni2001,anker2005}.
Controlled disordered potentials can also be produced optically
as demonstrated in several recent experiments
\cite{lye2005,clement2005,fort2005,schulte2005,clement2006},
for instance using speckle patterns \cite{horak1998,grynberg2000}.
Other techniques can be
employed to produce controlled disorder such as
the use of magnetic traps designed on atomic chips with rough
wires \cite{leanhardt2003,jones2004,kraft2002,wang2004,esteve2004},
the use of localized impurity atoms \cite{gavish2005,paredes2005},
or the use of radio-frequency fields \cite{courteille2006}.
However, the use of speckle potentials has unprecedented advantages
from both practical and fundamental points of view.
First, they are created using simple optical devices and their statistical
properties are very well known \cite{goodman1975,goodman2007}.
Second, they have finite-range correlations which offers richer situations than
theoretical $\delta$-correlated potentials (\ie uncorrelated disorder) and
the correlation functions can be designed almost at will by changing the geometry
of the optical devices \cite{goodman1975,goodman2007}.
Finally, both the amplitude and the correlation length (down to fractions of micrometers)
can be controlled accurately and calibrated using ultracold atoms \cite{clement2006}.

Within the context of ultracold gases, important theoretical efforts have been
devoted to disordered optical lattices which mimic the Hubbard model
\cite{fisher1989,jaksch1998,jaksch2005,greiner2002,lewenstein2007}. For bosons, quantum phase transitions from superfluid
to Bose glass and Mott insulator phases have been predicted \cite{roth2003,damski2003}
and evidence of the Bose glass has been obtained experimentally \cite{fallani2007}.
With Fermi-Bose mixtures, the phase diagram is even richer and includes
the formation of a Fermi-glass, a quantum percolating phase and a spin glass
\cite{sanpera2004,ahufinger2005,sanpera2004bis}.
Effects of disorder in Bose gases at equilibrium without optical lattice have been addressed
in connection with the behavior of the BEC phase transition \cite{lenoble2004,falco2007},
the quantum states of Bose gases \cite{martino2005,lsp2006,lugan2007a,yukalov2007},
the localization of Bogolyubov quasi-particles \cite{bilas2006,lugan2007b},
the dynamics of time-of-flight imaging of disordered BECs \cite{clement2007,chen2007},
and random-field-induced order in two-component Bose gases \cite{wehr2006,niederberger2007}.

	%%%%%%%%%%%%%%%%%%%%%%%%%%%%%%%%%%%
	\subsection{Scope and main results of the paper}
\label{introduction.outlook}

The dynamics of BECs in disordered (or quasi-disordered) potentials is also attracting significant attention
in a quest for observing Anderson localization in non-interacting BECs \cite{gavish2005,kuhn2005,kuhn2007}
or in BECs with repulsive interactions \cite{lsp2007,lsp2005,paul2005,paul2007}.
Recent experiments have demonstrated the strong suppression of transport in expanding BECs in the
presence of optical speckle potentials \cite{clement2005,fort2005,clement2006}, but this effect is
not related to Anderson localization \cite{clement2005}.

In this paper, we theoretically and numerically analyze the expansion of an interacting
one-dimensional (1D) BEC in a disordered potential.
We focus on a regime where the inter-atomic interactions {\it initially} exceed the
kinetic energy (Thomas-Fermi regime), a situation that significantly differs
from the textbook Anderson localization problem but which is relevant for almost all
current experiments with disordered BECs
\cite{clement2005,fort2005,schulte2005,clement2006,clement2007,chen2007}.
We distinguish two regimes that we name {\it strong disorder} and {\it weak disorder} respectively.

The case of strong disorder corresponds to the situation of the experiments of 
Refs.~\cite{clement2005,fort2005,clement2006} where
the interaction energy in the center of the BEC remains large during the expansion and
where the reflection coefficient
from a single modulation of the disordered potential is of the order of unity.
In this case, our numerical results reproduce the strong suppression of the transport of the BEC
as observed in the experiments of Refs.~\cite{clement2005,fort2005,clement2006}.
We analyze the scenario of {\it disorder-induced trapping} proposed in Ref.~\cite{clement2005}
in which two regions of the BEC are identified. The first region corresponds to the center,
where the trapping results from a competition between the interactions and the disorder.
The second region corresponds to the tails of the BEC where almost free particles
are multiply scattered from the modulations of the disordered potential.
There, localization is rather due to the competition between the kinetic energy 
and the disordered potential but is ultimately due to the almost total classical reflection
of the matterwave from a single barrier.
These  two effects are responsible for blurring Anderson localization effects \cite{clement2005,fort2005}.

Weak disorder corresponds to a situation where the probability of large and wide modulations
of the disordered potential is small. In this case, we show that {\it Anderson localization}
does occur
as a result of multiple quantum scattering from the modulations of the disordered potential.
Let us briefly describe the scenario first proposed in Ref.~\cite{lsp2007}.
Initially, the repulsive interactions are important as compared to the kinetic energy
and to the potential energy associated to the disordered potential. Then, the interactions induce the expansion
of the BEC and determine the momentum distribution of the BEC.
After a time typically equal to the inverse of the initial trapping frequency, the interactions
vanish and the momentum distribution reaches a steady state. 
Then, the BEC is a superposition of non-interacting waves of momentum $k$. Each wave localizes
with its own localization length $\Lloc (k)$.
By calculating analytically the superposition of the localized waves, we show that
the BEC can be exponentially localized or only show an algebraic decay depending
on the correlation function of the disordered potential.
In particular, due to peculiar long-range correlations, the BEC localizes exponentially
in speckle potentials only if $\xiini>\sigmar$, where $\xiini$
is the initial healing length of the BEC and $\sigmar$ is the correlation length of
the disorder.

	%%%%%%%%%%%%%%%%%%%%%%%%%%%%%%%%%%%
	\subsection{Organization of the paper}
\label{introduction.plan}

The paper is organized as follows.
In section~\ref{equilibrium}, we review the properties of a BEC at equilibrium
in a combined harmonic plus disordered potential, in
particular in the non-trivial regime where the healing length of the BEC exceeds
the correlation length of the disordered potential.
The next two sections deal with the expansion of an interacting BEC in a disordered potential.
Section~\ref{transport} is devoted to the case of strong disorder.
We reproduce and complete our previous results \cite{clement2005} which demonstrate
the suppression of the expansion of the BEC in
a speckle potential with similar parameters as in the experiments of
Refs.~\cite{clement2005,fort2005,clement2006}.
The scenario of disorder-induced trapping is analyzed and characteristic properties of
the BEC trapped by the disorder are calculated analytically and compared to numerical results.
In particular we derive an analytic expression for the central density of the BEC trapped by disorder
which happens to be characteristic of the disorder-induced trapping phenomenon and we show that the
ultimate suppression of the expansion of the BEC is due to classical reflections from
the large modulations of the disordered potential.
We also compare these findings with the case of a BEC expanding
in a periodic potential with similar characteristics as the disordered potential.
Section~\ref{anderson} is devoted to the case of weak disorder.
We show that Anderson localization
can show up in an expanding, interacting BEC under appropriate conditions that are clarified.
We show that the localization properties of the density profile crucially depend on
both the momentum distribution of the expanding BEC and
the correlation function of the disordered potential.
In particular, in the case of a speckle potential, we find a 1D {\it effective mobility edge}.
We calculate analytically the expected localization lengths
and compare our findings to the results of numerical calculations.
Finally in section~\ref{conclusion}, we summarize our findings and discuss expected impacts
of our work on experiments on disordered BECs.

%%%%%%%%%%%%%%%%%%%%%%%%%%%%%%%%%%%%%%%%%%%%%%%%%%%%%%%%%%%%%%%%%%%%%%
\section{Condensates at equilibrium in a combined harmonic trap plus disordered potential}
\label{equilibrium}

	%%%%%%%%%%%%%%%%%%%%%%%%%%%%%%%%%%%
	\subsection{Interacting Bose-Einstein condensates in a 1D inhomogeneous potential}
\label{equilibrium:system}

We consider a low-temperature 1D Bose gas with short-range atom-atom
interactions $\goned \delta(z)$ where $\goned$ is the 1D coupling constant.
The Bose gas is assumed to be subjected to (i) a harmonic potential of frequency
$\omega$ and (ii) an additional inhomogeneous potential $\Vopt (z)$.
In a finite system as considered in this work,
assuming weak interactions, \ie $\overline{n} \gg m \goned / \hbar^2$
where $\overline{n}$ is the average density and $m$ the atomic mass
\cite{olshanii1998,petrov2000}, the Bose gas will form a Bose-Einstein condensate
even in low-dimensional (\eg 1D) geometries \cite{petrov2000}.
Hence, we can treat the BEC within the mean-field approach \cite{dalfovo1999,pitaevskii2004} using the 
Gross-Pitaevskii equation (GPE):
%+++++++++++++++++++++++++++++++++++++++++%
\begin{eqnarray}
i\hbar \partial_t \psi (z,t) 
& = & \left[ \frac{-\hbar^2 \partial_z^2 }{2m} + \frac{m\omega^2 z^2}{2} + \Vopt (z) \right. \label{GPE} \\
&&           \hspace{1.cm} \left. + \goned |\psi (z,t)|^2 - \mu \phantom{\frac{0}{0}}\hspace{-0.2cm} \right] \psi (z,t), \nonumber 
\end{eqnarray}
%+++++++++++++++++++++++++++++++++++++++++% 
where $\mu$ is the BEC chemical potential.

In the following, we investigate the situations where the additional potential 
reads $\Vopt(z)=\Vr v(z)$ with $v(z)$ being either a disordered or a periodic 
function with vanishing average and unity standard deviation. Therefore, we have
$\langle \Vopt(z) \rangle = 0$ and
$\sqrt{\langle \Vopt(z)^2 \rangle - \langle \Vopt(z) \rangle^2} = |\Vr|$.
The sign of $\Vr$ depends on the definition of the function $v(z)$ and on the kind
of potential one considers. For instance, in optical speckle potentials, the quantity
$v(z)+1$ is defined to be positive and
$\Vr>0$ for blue-detuned laser light (case of the experiments of Refs.~\cite{clement2005,clement2006,clement2007})
while $\Vr<0$ for red-detuned laser light (case of the experiments of Refs.~\cite{lye2005,fort2005,chen2007}).
For a sine-periodic potential, using $\Vr<0$ or $\Vr>0$ does not change the physics.
See \ref{potentials} for details.

	%%%%%%%%%%%%%%%%%%%%%%%%%%%%%%%%%%%
	\subsection{The Bose-Einstein condensate wavefunction}
\label{equilibrium:bec}
Here, we briefly discuss the influence of an inhomogeneous potential on the BEC at equilibrium in
the harmonic trap. We assume that the amplitude of the disordered potential is smaller than the
chemical potential of the BEC ($\Vr \ll \mu$) and that $\mu \gg \hbar \omega$.
This question has been investigated in details in Ref.~\cite{lsp2006}. Here, we only outline the results.

At equilibrium, the BEC wavefunction is real (up to an irrelevant uniform phase) and is the solution
of Eq.~(\ref{GPE}) with $\partial_t\psi=0$:
%+++++++++++++++++++++++++++++++++++++++++%
\begin{equation}
\mu \psi (z) = \left[ \frac{-\hbar^2 \partial_z^2 }{2m} + \frac{m\omega^2 z^2}{2} + \Vopt (z)
+ \goned |\psi (z)|^2 \right] \psi (z).
\label{sGPE}
\end{equation}
%+++++++++++++++++++++++++++++++++++++++++%
For $\Vr=0$ and $\mu \gg \hbar \omega$, the kinetic term can be neglected (Thomas-Fermi regime \cite{pitaevskii2004})
and the BEC wavefunction is $\psi_0 (z) = \sqrt{n_0 (z)}$ with
%+++++++++++++++++++++++++++++++++++++++++%
\begin{equation}
n_0(z)=\frac{\mu -m\omega^2 z^2/2}{\goned}.
\label{TF0}
\end{equation}
%+++++++++++++++++++++++++++++++++++++++++%
for $z$ such that $\mu > m\omega^2 z^2/2$ and $n_0(z)=0$ elsewhere.
The density profile $n_0(z)$ is an inverted parabola
of length $\LTF=\sqrt{2\mu / m\omega^2}$ much larger than the healing length
$\xiini=\hbar/\sqrt{4m\mu}$ \cite{pitaevskii2004}
(notice that this definition is different from
the one of Ref.~\cite{lsp2006}, where we used $\xi=\sqrt{2}\xiini$).

In the presence of an inhomogeneous potential ($\Vr \neq 0$), the parabolic shape of the density profile
is perturbed.
In general, the kinetic term cannot be neglected any longer. 
In particular, when $\xiini \gtrsim \sigmar$, the short-range modulations of the potential $V(z)$ induce
short-range modulations of the BEC wavefunction which contribute significantly in Eq.~(\ref{sGPE})
through the kinetic term.
In order to take into account the effect of the inhomogeneous potential, we use a perturbative approach
along the lines of Ref.~\cite{lsp2006}: we write $\psi (z) = \psi_0 (z) + \delta\psi (z)$ with $\delta\psi \ll \psi_0$.
The first order term of the perturbation series of Eq.~(\ref{sGPE}) is governed by the equation
%+++++++++++++++++++++++++++++++++++++++++%
\begin{equation}
-(\xiini^0)^2 \partial_z^2 (\delta \psi) 
+ \delta \psi
= - \frac{\Vopt (z) \psi_0}{2\goned n_0},
\label{GPEoneorder}
\end{equation}
%+++++++++++++++++++++++++++++++++++++++++%
where $\xiini^0 = \xiini / \sqrt{1-(z/\LTF)^2}$ is the local healing length.
Since $\LTF \gg (\xiini,\sigmar)$, it is legitimate to use the local
density approximation (LDA) \cite{pitaevskii2004}, \ie in a region
smaller than $\LTF$, the quantity $n_0$ can be considered as uniform.
In this approximation,
the solution of Eq.~(\ref{GPEoneorder}) is easily found by turning to the Fourier space. We find
$\delta\psi (q) = - {\tVopt (q) \psi_0}/{2\goned n_0}$
where
%+++++++++++++++++++++++++++++++++++++++++%
\begin{equation}
\tVopt (q) = \frac{\Vopt (q)}{1+(q\xiini^0)^2}
\label{smooth}
\end{equation}
%+++++++++++++++++++++++++++++++++++++++++%
and finally,
%+++++++++++++++++++++++++++++++++++++++++%
\begin{equation}
\psi (z) \simeq \psi_0 (z) \left[1 - \frac{\tVopt (z)}{2\goned n_0 (z)}\right],
\label{dpsiLDA}
\end{equation}
%+++++++++++++++++++++++++++++++++++++++++%
or equivalently,
%+++++++++++++++++++++++++++++++++++++++++%
\begin{equation}
n (z) \simeq n_0 (z) - {\tVopt (z)}/{\goned}.
\label{dpsiLDAbis}
\end{equation}
%+++++++++++++++++++++++++++++++++++++++++%
This solution justifies {\it a posteriori} the use of a perturbative approach for $\tVr \ll \mu$,
where 
$\tVr=\sqrt{\langle \tVopt(z)^2 \rangle - \langle \tVopt(z) \rangle^2}$ is the standard
deviation of the potential $\tVopt (z)$. Notice that the equality $\langle \tVopt(z) \rangle = 0$
directly follows from Eq.~(\ref{smooth}).

An important consequence of the solutions~(\ref{dpsiLDA}),(\ref{dpsiLDAbis})
is that the BEC wavefunction is only
weakly perturbed by the inhomogeneous potential $V(z)$ if $\tVr \ll \mu$.
It follows from Eq.~(\ref{smooth}) that for $\xiini \ll \sigmar$,
$\tVopt (z) \simeq \Vopt (z)$
and the relative inhomogeneities of the BEC density are $\delta n/n \sim \Vr/\mu \ll 1$.
For $\xiini \gtrsim \sigmar$, the relative inhomogeneities are even smaller since
all Fourier components of $\tVopt$ are smaller than those of $\Vopt$.
More precisely, the effective potential $\tVopt$ is roughly obtained from $\Vopt$ by
suppressing the Fourier components with a wavelength smaller than the healing length.
In other words, the BEC density does not follow the modulations of the
bare disordered potential $\Vopt (z)$ but actually follows the smoother modulations of
the {\it smoothed disordered potential} $\tVopt (z)$.

Therefore, an interacting BEC {\it at equilibrium} in a disordered potential is not
localized in the sense of Anderson.
One may wonder whether this conclusion still holds for stronger disorder or weaker interactions,
where the meanfield approach can break down.
This question has been addressed in Ref.~\cite{lugan2007a}. It turns out that for very weak interactions,
the Bose gas forms a so-called {\it Lifshits glass} which corresponds to a Fock state of various
localized single-particle states. These states belong to the Lifshits tail of the non-interacting
spectrum and are strongly trapped. Therefore, Anderson localization can hardly be observed
unambiguously in this case.
It seems more favorable to find evidences of Anderson localization in transport experiments
of interacting BECs, rather than studying BECs at equilibrium in a disordered trap.

%%%%%%%%%%%%%%%%%%%%%%%%%%%%%%%%%%%%%%%%%%%%%%%%%%%%%%%%%%%%%%%%%%%%%%
\section{Strong disorder: Suppression of the expansion of a Bose-Einstein condensate in a
speckle potential and disorder-induced trapping scenario}
\label{transport}

In this section, we investigate the transport properties of a coherent BEC in a
disordered potential in the situation of the experiments of Refs.~\cite{clement2005,fort2005,schulte2005,clement2006}.
We thus assume
(i) that the chemical potential of the BEC is larger than the depth of the disordered potential
($\mu > \Vr$)
and (ii) that the correlation length of the disordered potential
is much larger than the healing length of the BEC and much smaller than the (initial)
size of the BEC, $\xiini \ll \sigmar \ll \LTF$.
We present numerical results which reproduce the suppression of the
transport of the BEC in a speckle potential, observed in Refs.~\cite{clement2005,fort2005,clement2006},
and discuss a scenario to explain this phenomenon.
In addition, we compare the observed behavior to the case of a periodic potential with
similar characteristics.

	%%%%%%%%%%%%%%%%%%%%%%%%%%%%%%%%%%%
	\subsection{Expansion of an interacting BEC in a speckle potential}
\label{transport:expansion}
In order to induce transport, we start from a BEC at equilibrium in the
harmonic and disordered potentials (see Sec.~\ref{equilibrium}).
At time $t=0$, we suddenly switch off the trapping harmonic potential,
keeping the disordered potential. This process is similar to the one used in
Refs~\cite{clement2005,fort2005,clement2006,lsp2005}.
The evolution of the BEC is thus governed by the GPE~(\ref{GPE}) with $\omega=0$ and 
the initial condition corresponds to the TF wavefunction discussed in Sec.~\ref{equilibrium:bec}.

\begin{figure}[t!]
\begin{center}
\infig{27.em}{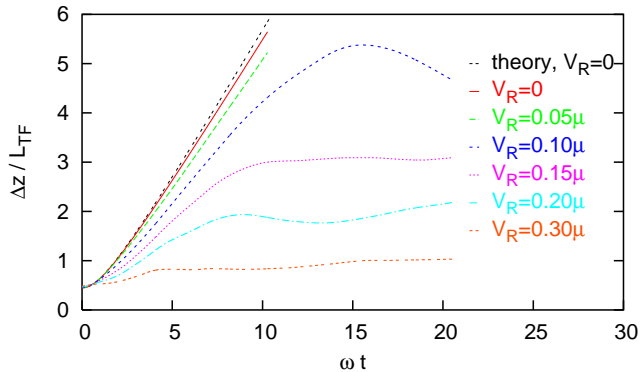}
\end{center}
\caption{(color online)
Time-evolution of the rms-size of the BEC wavefunction evolving in the
disordered potential $\Vopt$ for several values of the amplitude $\Vr$. 
The (black) dashed line is the theoretical prediction of the scaling
theory~(\ref{scalingparam}) with a vanishing disordered potential.
Here, we have used $\sigmar=0.012 \LTF$ and $\xiini = 5.7\times 10^{-4} \LTF$.
} 
\label{expan}
\end{figure}

The time evolution of the {\it root mean square} (rms) size of the BEC,
$\Delta z (t) = \sqrt{\langle z^2 \rangle - \langle z \rangle^2}$,
as obtained from the numerical integration of the time-dependent GPE~(\ref{GPE}),
is plotted in Fig.~\ref{expan} for several amplitudes $\Vr$ of the disordered potential.
In the absence of disorder, the interacting BEC expands self-similarly
as predicted by the scaling approach \cite{kagan1996,castin1996}:
%+++++++++++++++++++++++++++++++++++++++++%
\begin{equation}
\psi (z,t) \simeq 
\frac{\psi [z/b(t),0]}{\sqrt{b(t)}}
\exp\left( i \frac{m z^2 \dot{b}(t)}{2\hbar b(t)} \right)
\label{scalingwf}
\end{equation}
%+++++++++++++++++++++++++++++++++++++++++%
where $b(t)$ is the {\it scaling parameter} which is governed by the equation
%+++++++++++++++++++++++++++++++++++++++++%
\begin{equation}
\ddot{b} (t)=\omega^2/b^2(t)
\label{scaling1}
\end{equation}
%+++++++++++++++++++++++++++++++++++++++++%
with the initial conditions $b(t=0)=1$ and $\dot{b}(t=0)=0$.
Integrating these equations, we find
%+++++++++++++++++++++++++++++++++++++++++%
\begin{equation}
\sqrt{b(t) (b(t)-1)} + \ln [\sqrt{b(t)}+\sqrt{b(t)-1}] = \sqrt{2} \omega t,
\label{scalingparam}
\end{equation}
%+++++++++++++++++++++++++++++++++++++++++%
which asymptotically reduces to a linear expansion at large time, $b(t) \sim \sqrt{2} \omega t$.
The numerical calculations agree with this expression as shown in Fig.~\ref{expan}.

%-----------------------------------------%
%-----------------------------------------%

The situation is significantly different in the presence of disorder.
In this case, the initial
BEC wavefunction is the usual Thomas-Fermi inverted parabola perturbed by the disordered potential 
\cite{lsp2006}.
For $t \lesssim 1/\omega$, the scaling form~(\ref{scalingwf}) is still a good solution of the GPE 
and, according to the scaling theory \cite{kagan1996,castin1996} the BEC wavefunction expands.
For larger times and small amplitudes of the disordered potential 
($\Vr \lesssim 0.1 \mu$), the effect of disorder on the expansion observed in the numerical calculations
is small and the BEC expands by about one order of magnitude for $\omega \tau = 10$.
For larger amplitudes of the disorder ($\Vr \gtrsim 0.15 \mu$),
the expansion of the BEC stops after an initial expansion stage described above.
This effect signals the localization of the BEC wavefunction due to the presence of disorder.

\begin{figure}[t!]
\begin{center}
\infig{27.em}{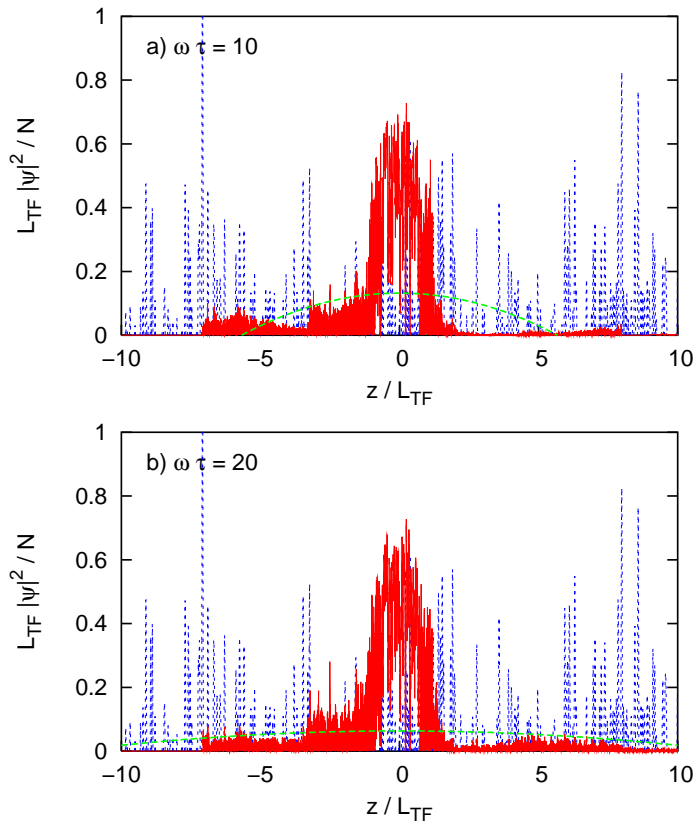}
\end{center}
\caption{(color online)
Density profiles of the BEC for $\Vr=0.2 \mu$, $\sigmar=0.012 \LTF$ 
and $\xiini = 5.7\times 10^{-4} \LTF$ (solid red lines)
for two different values of the expansion time $\tau$ and expected Thomas-Fermi
profiles in the  absence of a disordered potential (dashed green lines).
We also show the disordered potential normalized so as to be homogeneous
to a density ($\Vopt/\goned$; dotted blue line).
} 
\label{density}
\end{figure}

Important information can be obtained from density profiles of the localized
BEC. For instance, density profiles corresponding to a single evolution are plotted at two different times in Fig.~\ref{density}.
From these, it appears that the localized BEC is made of
two distinct parts: a static dense core and fluctuating dilute tails (see also Fig.\ref{scaledensity}).
In particular, the small fluctuations of $\Delta z$ observed in Fig.~\ref{expan} are due to the contribution
of the tails of the BEC that still evolve while the core of the wavefunction is localized
(see section~\ref{transport:scenario}).

%-----------------------------------------%
%-----------------------------------------%

It is worth noticing that the BEC expansion stops for amplitudes of disorder 
significantly smaller than the typical energy per particle in the initial BEC: $\Vr < \mu$.
This {\it suppression of transport} is phenomenologically similar to what is expected
from Anderson localization \cite{anderson1958,nagaoka1982,ando1988}.
Strictly speaking, Anderson localization relies on the existence of
localized {\it single-particle eigenstates} and on the subsequent absence of
diffusion \cite{anderson1958}.
However, we have stressed that the presence of predominant inter-atomic interactions dramatically
changes the picture \cite{clement2005}.
On one hand, repulsive interactions are expected to reduce the localization effect
\cite{lsp2006,shepelyanski1994}.
During the initial expansion of the BEC, the interaction energy greatly 
dominates over the kinetic energy in the center of the BEC so that no Anderson-like localization effect
is expected in this region.
On the other hand, although the particles in the tails are weakly interacting due to the small density,
the initial interactions determine their typical energy as the initial expansion stage converts the
interaction energy into kinetic energy.
We will see that, for strong disorder as considered here, the modulations of the disordered 
potential will ultimately stop the expansion of the dilute tails, masking any Anderson-like effect
(in the case of weak disorder however, Anderson localization can be obtained in this region as discussed in Sec.~\ref{anderson}).
In the following, we detail the scenario of {\it disorder-induced trapping} outlined
above and first proposed in Ref.~\cite{clement2005}.

	%%%%%%%%%%%%%%%%%%%%%%%%%%%%%%%%%%%
	\subsection{Scenario of disorder-induced trapping}
\label{transport:scenario}
The dynamics of the BEC in the disordered potential is governed by three different forms of energy: 
(i) the potential energy associated to the disordered potential,
(ii) the interaction energy and 
(iii) the kinetic energy.
It is thus useful to evaluate and compare the kinetic and interaction
energies to understand the behavior of the BEC in the disordered potential.
To this end, notice first that it follows from the initial expansion
of the BEC that the fast atoms populate the tails of the expanding BEC
while the slow atoms stay close to the center.
In addition, notice that, except for very small amplitudes of the disordered
potential and subsequent long expansion times, the density in the core of the
BEC remains large whereas it drops to zero in the tails (see Fig.~\ref{density}).
We thus distinguish two different regions of the BEC:
(i) the core where the density is large and the interaction energy is dominant and 
(ii) the tails where the density is small and the kinetic energy dominates.
The behavior of the BEC turns out to be completely different in
these two regions \cite{clement2005}.

\subsubsection{Quasi-static Thomas-Fermi profile in the core of the BEC}~-~
\label{center}
For the sake of clarity, we define the core of the BEC as half the total size of the initial condensate:
$-\LTF/2 < z < \LTF/2$ and call
%+++++++++++++++++++++++++++++++++++++++++%
\begin{equation}
\nc(t) = \frac{1}{\LTF}
		  \int_{-\LTF/2}^{+\LTF/2} 
		  dz\ |\psi(z,t)|^2,
\label{n0}
\end{equation}
%+++++++++++++++++++++++++++++++++++++++++%
the average BEC density in the center. In particular, at time $t=0$, due to the parabolic envelope
resulting from the harmonic trap, we find
%+++++++++++++++++++++++++++++++++++++++++%
\begin{equation}
\nc(t=0) = \frac{11}{12}\frac{\mu}{\goned}
\label{n0init}
\end{equation}
%+++++++++++++++++++++++++++++++++++++++++%
in the absence of disorder but also in the presence of a self-averaging disordered
potential\footnote{In the context of disordered systems, a quantity is said to be
`self-averaging' when it verifies the principle of `spatial ergodicity'.
In other words, it means that
the average over realizations of the disordered potentials of a relevant quantity, F,
equals the corresponding spatial average:
\ie $\langle F \rangle \simeq \frac{1}{L}\int_0^L dz F(z)$.}.

During the initial expansion stage, the average density in the core,
$\nc$, slowly decreases and the parabolic envelope disappears.
Since the interaction energy
significantly exceeds the kinetic energy, we expect the local density $|\psi(z,t)|^2$ 
to follow almost adiabatically the instantaneous
value of $\nc(t)$ approximately in the Thomas-Fermi regime
so that
%+++++++++++++++++++++++++++++++++++++++++%
\begin{equation}
|\psi (z,t)|^2 \simeq \nc(t) - \Vopt(z) / \goned.
\label{TFbis}
\end{equation}
%+++++++++++++++++++++++++++++++++++++++++%

\begin{figure}[t!]
\begin{center}
\includegraphics[width=8.5cm]{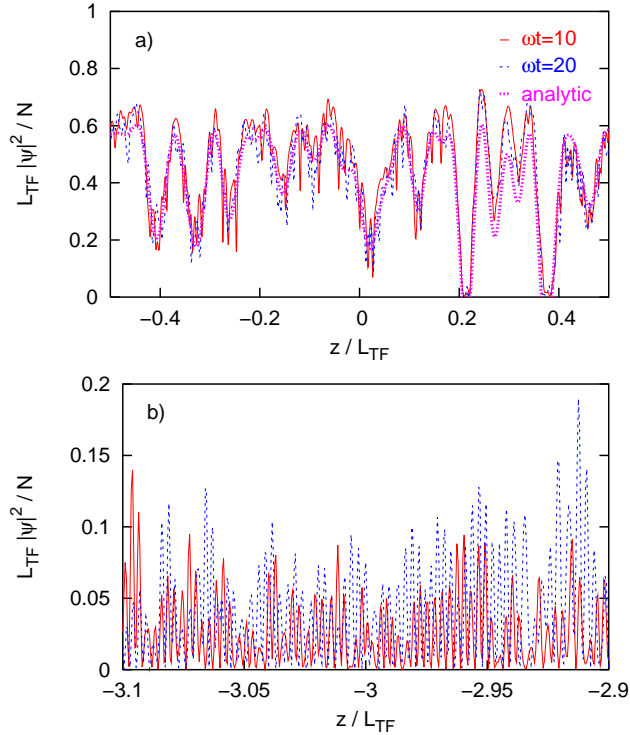}
\end{center}
\caption{(color online)
Density profiles of the BEC during the evolution in the disordered
potential at different times for $\Vr=0.2\mu$ in the core (a)
and in the tails (b) of the BEC.
Both are magnifications of the plots of Fig.~\ref{density}.
The solid (red online) and dashed (blue online) lines correspond 
respectively to the times $\omega t=10$ and $\omega t=20$
of the same evolution
and the dotted (purple online) line corresponds to Eq.~(\ref{TFbis}) with $\nc$ as 
a fitting parameter.
Notice the different scales in the two figures.}
\label{scaledensity}
\end{figure}

In order to check this prediction, we plot in Fig.~\ref{scaledensity}a
the result of the numerical integration of the GPE~(\ref{GPE}) for
the density profile in the central region of the BEC during the evolution
in the disordered potential at two different times, together with a plot of the
analytical expression~(\ref{TFbis}).
In particular, two properties are of special interest here.
First, we observe that the time-dependent fluctuations of
the density profile are significantly smaller than the
modulations of the disordered potential $\Vopt (z)/\goned$.
Second, the density profiles are in good agreement with Eq.~(\ref{TFbis}).
This observation supports the scenario of an adiabatic decrease of the density in the center of
the BEC.
The value of $\nc$ at the end of the expansion turns out to be characteristic
of this scenario.
In the following, we show that $\nc$ can indeed be computed from the statistical
properties of the disordered potential.

%-----------------------------------------%
%-----------------------------------------%

The expansion of the core of the BEC in the disordered potential stops
when the condensate {\it fragments}, \ie when the effective chemical
potential in the center of the BEC ($\overline{\mu}=\goned \nc$) decreases down to 
the value of typically two large modulations of the disordered potential. 
At this time, the energy per particle in the core of the BEC becomes too small to
over-pass the potential barriers and the core of the BEC gets trapped
between these large modulations. This scenario allows us to
determine the final value of the average density $\nc$ in the core of 
the BEC.
Let us call $\Npeaks(V)$ the number of maxima of the disordered potential in 
the central part of the BEC ($-\LTF/2<z<\LTF/2$) 
with an amplitude larger than a given value $V$ and
assume that it can be computed from the statistical
properties of the disordered potential.
The density in the center of the BEC after the trapping has occurred thus corresponds to the
maximum value of $\nc$ below which two modulations of $\Vopt$ in average
are present in the center of the BEC.
This is simply computed by solving for $\Npeaks (V=\nc \goned)=2$.
Although, this scheme is general, it appears clearer when applied to a case
where $\Npeaks(V)$ can be explicitly computed.
Let us now consider the case of a speckle potential \cite{goodman1975,goodman2007}
with $\Vr > 0$.
It is shown in \ref{potentials} [see Eq.~(\ref{picseqtext})] that in this case 
$\Npeaks (V) \simeq \alpha \left(\frac{\LTF}{\sigmar}\right) \exp\left[ -\beta \frac{V}{\Vr} \right]$ where $\alpha \simeq 0.30$ and $\beta \simeq 0.75$.
From this, we easily find that the final density of the core of the BEC is
$\nc \simeq \frac{1}{\beta} \left(\frac{\Vr}{\goned}\right) \ln \left[ \frac{\alpha \LTF}{2\sigmar} \right]$.
In addition, we notice that the final density cannot exceed the initial density
as resulting from an expansion. Therefore Eq.~(\ref{n0eq}) is valid
only for $\frac{1}{\beta} \left(\frac{\Vr}{\goned}\right) \ln \left[ \frac{\alpha \LTF}{2\sigmar} \right] \lesssim \mu/\goned$. In the opposite
situation, the BEC is already multiply fragmented at $t=0$ and
the final density saturates at
$\nc \simeq \frac{11}{12} \frac{\mu}{\goned}$
[see Eq.~(\ref{n0init})].
In summary, we expect that the average density of the BEC trapped by the disorder is
%+++++++++++++++++++++++++++++++++++++++++%
\begin{equation}
\nc \simeq \min \left\{\frac{1}{\beta} \left(\frac{\Vr}{\goned}\right) 
	               \ln \left[ 
		       \frac{\alpha \LTF}{2\sigmar} \right],
		       \frac{11\mu}{12\goned}\right\}.
\label{n0eq}
\end{equation}
%+++++++++++++++++++++++++++++++++++++++++%

\begin{figure}[t!]
\begin{center}
\includegraphics[width=8.5cm]{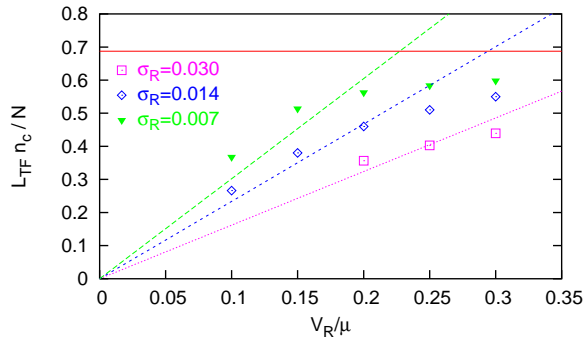}
\end{center}
\caption{(color online)
Average density $\nc$ in the core of the BEC trapped by the disorder versus
the amplitude of the disordered potential $\Vr$ for different values of the correlation
length $\sigmar$ and comparison to Eq.~(\ref{n0eq}). The horizontal (red online) line
corresponds to the saturation limit $\nc=11\mu / 12\goned$.
} 
\label{fign0}
\end{figure}

In order to check Eq.~(\ref{n0eq}), we have  extracted the averaged central density [see Eq.~(\ref{n0})]
from
the wavefunctions $\psi$ calculated numerically for several amplitudes $\Vr$ and correlation lengths 
$\sigmar$ of the disordered potential. 
In Fig.~\ref{fign0}, we plot $\nc$ as a function of $\Vr$ for several 
$\sigmar$ together with the prediction~(\ref{n0eq}).
The results show that Eq.~(\ref{n0eq}) provides a good estimate
of the final density $\nc$ in the core of the BEC. In particular, for
small amplitudes of the disorder, $\nc$ grows linearly with $\Vr$ with a 
coefficient in agreement with Eq.~(\ref{n0eq}) up to about $10~\%$. 
For larger amplitudes of the disorder, $\nc$ saturates below $11\mu / 12\goned$
as expected.

%-----------------------------------------%
%-----------------------------------------%

This behavior agrees with experimental results for a blue-detuned speckle
potential ($\Vr>0$) \cite{clement2006}.
It is worth noticing that our scenario is expected to apply also to the case of a
red-detuned speckle potential ($\Vr < 0$) as used in Ref.~\cite{fort2005}.
In this case, the fragmentation occurs when $\overline{\mu}=|\Vr|$ 
independently of the correlation length of disorder (if $\xiini \ll \sigmar$). 
Then, the fragmented 
BEC is trapped in the small wells of the disordered potential with a typical size
$\sigmar$ and with a central density $\nc \simeq |\Vr|/\goned$
(independent of $\sigmar$).
Instead, for a blue-detuned speckle potential as investigated above, the BEC is
trapped between large modulations that may be separated by a distance much larger
than $\sigmar$. As a consequence the final density at the center is expected to be
significantly larger. This is confirmed by the numerical results of Ref.~\cite{modugno2005}.

\subsubsection{Strong reflections in the tails of the BEC}~-~
\label{tails}
The situation is completely different in the tails of the BEC.
Due to the small atomic density, the kinetic energy now dominates over the interaction energy.
The tails are populated by fast moving, weakly interacting atoms that
undergo multiple scattering from the modulations of the
disordered potential. Ultimately, the trapping of these atoms results
from almost total classical reflection on a single large modulation of the disordered
potential with an amplitude exceeding the typical energy of a single particle
\cite{clement2005,fort2005}.
This scenario is supported by the density profiles plotted in 
Fig.~\ref{density} where one can observe a sharp drop of 
the atomic density at the edges of the BEC (\ie at positions 
$z_\textrm{min} \simeq -7 \LTF$ and 
$z_\textrm{max} \simeq 8 \LTF$ in Fig.~\ref{density}).
Notice that significant drops correspond either 
(i) to modulations of the disordered potential larger than the initial chemical 
potential $\mu$ (\eg at $z_\textrm{min} \simeq -7 \LTF$) or 
(ii) to a concentration of weaker barriers (\eg at 
$z_\textrm{min} \simeq -3.5 \LTF$).

In contrast with the situation in the core of the BEC, it is expected that
(i) the density profile does not show a Thomas-Fermi shape and 
(ii) the local density is not stationary.
Both properties agree with our numerical results as shown
in Fig.~\ref{scaledensity}b where we plot a magnification of a small region
corresponding to the tails of the BEC of Fig.~\ref{density}.
In particular, the shorter modulations of the wavefunction observed
in Fig.~\ref{scaledensity}b are due to the kinetic energy of the particles
in the tails.
This statement is corroborated by the calculation of the energy per particle,
$\epsilon$.
Due to energy conservation, the energy can be computed at the initial time $t=0$
(\ie right after releasing the BEC from the trapping potential),
$\epsilon = \frac{1}{N} \int d z \ |\psi(z)|^2
\left[ \Vopt (z) + \goned |\psi(z)|^2/2 \right]$.
Using Eq.~(\ref{dpsiLDAbis}), we easily find that
%+++++++++++++++++++++++++++++++++++++++++%
\begin{equation}
\epsilon = \frac{2\mu}{5} \left[1-\frac{15}{8}\left(\frac{\Vr}{\mu}\right)^2\right].
\label{Etotbis}
\end{equation}
%+++++++++++++++++++++++++++++++++++++++++%
The disordered potential perturbs the energy per particle only at second order in $\Vr/\mu$,
and, for $\Vr \ll \mu$, we have $\epsilon \propto \mu$.
From this, we expect that the typical wavelength $\Lambda$ of the fluctuations 
in the tails would be of the order of the healing length in the initial 
condensate, so that
$ \Lambda / 2\pi \sim \xiini$.
This is confirmed by the properties of the momentum distribution
of the BEC which show two sharp peaks located around
$p \simeq \pm \hbar/\xiini$.

	%%%%%%%%%%%%%%%%%%%%%%%%%%%%%%%%%%%
	\subsection{Expansion of a condensate in a periodic potential}
	\label{periodic}
Up to this point, our analysis has shown how the competition between inter-atomic
interactions and disorder (in the center of the BEC) or
between kinetic energy and disorder (in the tails of the BEC) can strongly
suppress the coherent transport of an interacting matterwave in a disordered potential.
A natural extension of our analysis is to compare these findings
to the situation in a periodic potential with similar characteristics (see
\ref{potentials}).
In the case of a periodic potential, no suppression of transport is expected as no
large peak can provide a sharp stopping of the expansion,
and obviously, no `\`a la Anderson' localization should occur.

\begin{figure}[t!]
\begin{center}
\infig{27.em}{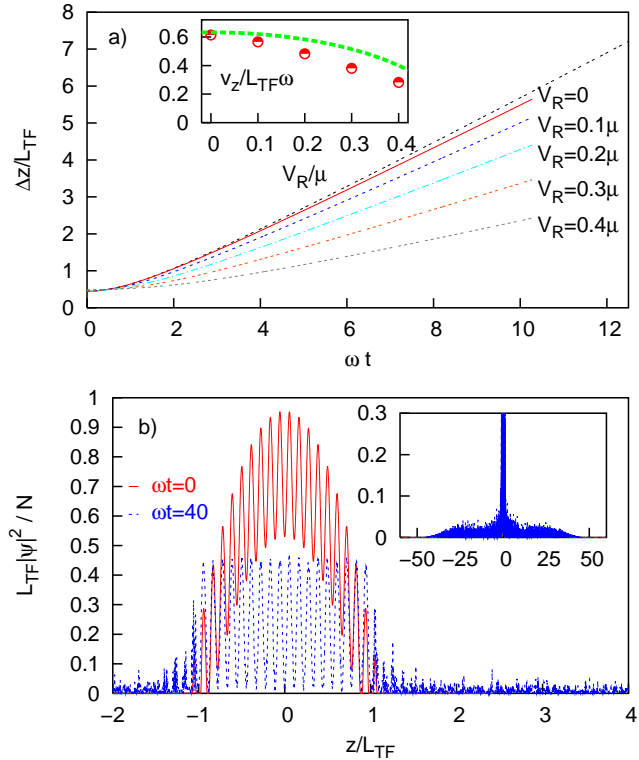}
\end{center}
\caption{(color online)
a) Time-evolution of the rms-size of the BEC wavefunction evolving in 
a periodic potential for several amplitudes $\Vr$ and for $\lambda=0.11\LTF$.
The theoretical prediction corresponding to Eq.~(\ref{scalingparam}) in free space is also shown
(black dotted line).
The inset shows the velocity of the expansion of the BEC together with the 
theoretical estimate~(\ref{veloth}).
b) Density profiles of the BEC in the harmonic trap and after an expansion time in the 
periodic potential of $t=40/\omega$ for $\Vr=0.2\mu$.
} 
\label{expanperio}
\end{figure}

Numerical results for the expansion of the BEC in the periodic potential described
in~\ref{potentials:periodic} are shown in Fig.~\ref{expanperio}a. The difference with the case
of a disordered potential (see Fig.~\ref{expan}) is striking:
as expected,
the BEC now expands linearly with time with an asymptotic expansion rate that decreases 
when the amplitude $\Vr$ of the periodic potential increases.
A detailed analysis shows however that the transport of the BEC in a periodic
potential and in a disordered potential share some properties for the parameters
used in this section, as we discuss below.

%-----------------------------------------%
%-----------------------------------------%

\begin{figure}[t!]
\begin{center}
\includegraphics[width=8.5cm]{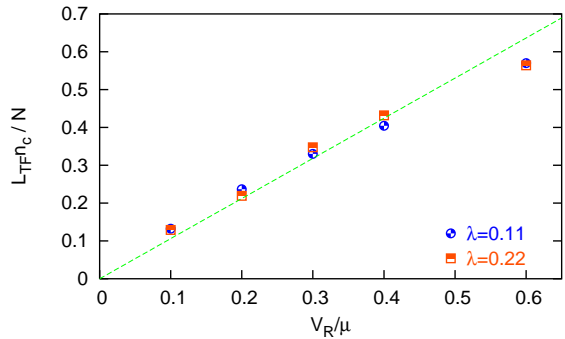}
\end{center}
\caption{(color online)
Average density in the center of the BEC trapped in the periodic potential
versus the lattice depth $\Vr$ and for two lattice spacings $\lambda$.
The theoretical prediction~(\ref{n0perioeq}) is also plotted (green dashed line).} 
\label{n0perio}
\end{figure}

Again, important information is contained in the density profiles such as the ones
plotted in Fig.~\ref{expanperio}b (to be compared to Fig.~\ref{density} which corresponds to
the disordered case).
Initially ($\omega t=0$),
the density profile follows the modulations of the periodic
potential modulated by the parabolic envelope associated to the harmonic
trapping \cite{lsp2006}.
During the initial expansion stage, the density in the center decreases slowly
and follows adiabatically a Thomas-Fermi shape with a slowly decreasing
instantaneous chemical potential $\overline{\mu}$.
Then, the evolution of the center stops when the chemical potential $\overline{\mu}$ 
exactly matches the potential depth, \ie when the BEC fragments.
This is similar to the disordered case.
However, in the case of a periodic potential, it is a deterministic process which
appears when
%+++++++++++++++++++++++++++++++++++++++++%
\begin{equation}
\nc \simeq \min \left\{ \sqrt{2} \left(\frac{\Vr}{\goned}\right), \frac{11\mu}{12\goned}
\right\}.
\label{n0perioeq}
\end{equation}
%+++++++++++++++++++++++++++++++++++++++++%
As shown in Fig.~\ref{n0perio}, this formula provides a very good value of 
the average density in the center of the BEC trapped in the periodic potential.
Remarkably, Eq.~(\ref{n0perioeq}) shows that $\nc$ does not depend on the
lattice spacing $\lambda$ as confirmed by the numerical results shown in
Fig.~\ref{n0perio}.
This is different from the case of a blue-detuned speckle potential ($\Vr>0$) where $\nc$ has been
shown to depend explicitly on the correlation length of
the disordered potential [see Eq.~(\ref{n0eq}) and Fig.~\ref{fign0}].

%-----------------------------------------%
%-----------------------------------------%

During the
subsequent evolution, a part of the BEC is thus trapped in the center while
the tails still expand as shown in Fig.~\ref{expanperio}.
Let us focus now onto the tails of the BEC. To do so,
let us first write the BEC wavefunction $\psi = \psi_\textrm{c}+\psi_\textrm{w}$
where $\psi_\textrm{c}$ and $\psi_\textrm{w}$ account for the center 
($|z|<\LTF$) and for the tails ($|z|>\LTF$) respectively.
As the supports of $\psi_\textrm{c}$ and $\psi_\textrm{w}$ are spatially
separated, we have
$\Delta z^2 = \int dz\ z^2 |\psi_\textrm{c}|^2 + \int dz\ z^2 |\psi_\textrm{w}|^2$.
The center gets trapped after a transient time so that 
$\int dz\ z^2 |\psi_\textrm{c}|^2$ tends to a 
constant, $\Delta z_ 0^2$, at large times. 
In contrast, the tails expand so that their density decrease.
After a typical time $~1/\omega$, a substantial part of the interaction energy is
converted into kinetic energy and the interaction term can be neglected for
the subsequent dynamics.
We also neglect the periodic potential which has a small amplitude compared to the
typical energy per particle in the tails.
Now, in free space,
$\Delta z^2_\textrm{w} = \frac{1}{N} \int dz\ z^2 |\psi_\textrm{w}|^2
\simeq \frac{2E_\textrm{w}}{Nm} t^2$
at large times where $E_\textrm{w}$ is the total (kinetic) energy in the tails
of the BEC. Due to the conservation of the total energy during the expansion,
we have $N\epsilon=E_\textrm{c}+E_\textrm{w}$ where $\epsilon$ is the energy
per particle given by
Eq.~(\ref{Etotbis}), and the energy in the center of the BEC,
$E_\textrm{c}=\Vr^2 \LTF/\goned$, is
easily computed from the Thomas-Fermi profile in the center of the BEC
[see Eq.~(\ref{TFbis})].
We finally find that 
$E_\textrm{w}/N \simeq \frac{2\mu}{5}\left[1-\frac{15}{4}\left(\frac{\Vr}{\mu}\right)^2\right]$
so that $\Delta z ^2 \simeq \Delta z_0 ^2 + v_z^2 t^2$ at times larger than
$1/\omega$ with
%+++++++++++++++++++++++++++++++++++++++++%
\begin{equation}
v_z \simeq \sqrt{\frac{2}{5}} \omega \LTF 
\left[ 1 - \frac{15}{4} \left(\frac{\Vr}{\mu}\right)^2\right]^{1/2}.
\label{veloth}
\end{equation}
%+++++++++++++++++++++++++++++++++++++++++%
In the absence of disorder ($\Vr=0$), Eq.~(\ref{veloth}) is consistent with the scaling
theory [Eq.~(\ref{scalingparam})] for which $\Delta z(t)=b(t)\LTF/\sqrt{5}$.
In the presence of disorder, it provides a reasonable agreement with the numerical findings
as shown in the Inset of Fig.~\ref{expanperio}a.
We attribute the discrepancy at the largest values of $\Vr$ to the main two
approximations that have been used.
First, a strict separation between the tails and the center
at $z=\LTF$ has been used to compute $E_\textrm{w}$ and we have thus neglected the
small intermediate region. Second, the interaction of the atoms with the
periodic potential is expected to increase the inertia of the expanding gas
and this should contribute to slightly lower the expansion velocity compared to the 
prediction~(\ref{veloth}).

%%%%%%%%%%%%%%%%%%%%%%%%%%%%%%%%%%%%%%%%%%%%%%%%%%%%%%%%%%%%%%%%%%%%%%
\section{Weak disorder: Onset of Anderson localization in the expansion of a condensate}
\label{anderson}

In section~\ref{transport}, we have shown that in the experimental conditions of
Refs.~\cite{clement2005,fort2005,clement2006}, Anderson localization effects are blurred
in an expanding, interacting BEC owing
(i) to important repulsive interactions in the center of the BEC, and
(ii) to strong reflections from single barriers of the disordered potential in the tails of the BEC.
Both effects are related to the presence of large modulations of the disordered potential.
It thus appears necessary to work in a parameter range where the probability of single
large modulations of the disorder is negligible in order to observe unambiguous
Anderson localization of an expanding BEC.

In this section, we work within the regime of weak disorder [a precise definition is given below,
see Eq.~(\ref{eq:lyapunovcond})].
Following the theory of Ref.~\cite{lsp2007}, we show that in this situation, Anderson localization of an
interacting BEC can be observed under appropriate conditions that we identify precisely \cite{lsp2007}.
We consider both cases of an impurity model of disorder and of a speckle potential.
In particular, for a speckle potential, we show that the long-range correlations induce
a 1D {\it effective mobility edge}, \ie strong exponential localization is obtained only
for $\xiini>\sigmar$.

	%%%%%%%%%%%%%%%%%%%%%%%%%%%%%%%%%%%
	\subsection{General model of Anderson localization of an expanding Bose-Einstein condensate}
\label{anderson:expansion}

\begin{figure}[t!]
\begin{center}
\includegraphics[width=8.5cm]{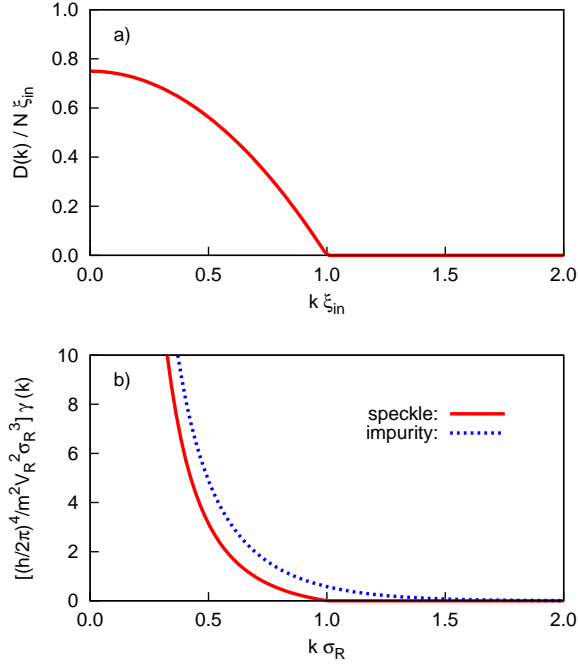}
\end{center}
\caption{(color online)
a) Stationary momentum distribution of an expanding, interacting Bose-Einstein condensate at large times ($t>1/\omega$).
b) Lyapunov exponent of single-particles of energy $E$ as a function of the momentum $k=\sqrt{2mE}/\hbar$.} 
\label{distgamma}
\end{figure}

\paragraph{Expansion of a Bose-Einstein condensate~- }
Let us examine again the expansion of the BEC in the disordered potential (see section~\ref{transport:expansion}).
For weak disorder, the initial interaction energy strongly
exceeds the potential energy associated with the disorder so that the first stage of expansion of the BEC is
hardly affected by the disorder.
For instance, the numerical results of Fig.~\ref{expan} for $\Vr=0.5\mu$ and $\Vr=0.1\mu$ confirm this
assertion for durations of expansion up to about $t \simeq 10/\omega$.
Within this time window, the momentum distribution
of the expanding BEC can thus be approximated to that of a BEC expanding in free space [see Eq.~(\ref{scalingwf})].
Calculating the Fourier transform of the scaling solution for interacting BECs
expanding in free space, $\psi (z,t)$, using the stationary phase approximation (valid for $t \gg \hbar/\mu$),
we find the momentum distribution
%+++++++++++++++++++++++++++++++++++++++++%
\begin{eqnarray}
\mathcal{D}(k,t) \simeq
\frac{3N \xiini / 4}{\sqrt{1-1/b(t)}}
&& \times 
\left[1 - \left( \frac{ k\xiini}{\sqrt{1-1/b(t)}} \right)^2 \right]
\nonumber \\
&& \times \Theta
\left[1 - \left( \frac{ k\xiini}{\sqrt{1-1/b(t)}} \right)^2 \right]
\label{momentth}
\end{eqnarray}
%+++++++++++++++++++++++++++++++++++++++++%
where $\Theta$ is the Heaviside step function.
Since $b(t) \simeq \sqrt{2}\omega t$, the momentum distribution reaches a steady-state at
times $t \gg 1/\omega$:
%+++++++++++++++++++++++++++++++++++++++++%
\begin{equation}
\mathcal{D}(k) \simeq
\frac{3N \xiini}{4}
\times
\left[1 - (k \xiini)^2 \right]
\times \Theta
\left[1 - (k \xiini)^2 \right].
\label{momentthstat}
\end{equation}
%+++++++++++++++++++++++++++++++++++++++++%
An important feature of the momentum distribution~(\ref{momentthstat}) is that it has a high-momentum
cut-off at $\kc=1/\xiini$ (see Fig.~\ref{distgamma}a).

For $t \gg 1/\omega$, almost all the initial interaction energy is converted into kinetic energy.
Neglecting the effect of disorder at this stage, this property can be obtained from the
scaling solution~(\ref{scalingwf}) \cite{kagan1996,castin1996}.
We find that the interaction energy is $\Eint (t) \simeq \Eint (0) / b(t)$.
Then using the property of conservation of the total energy during the expansion,
we find that the ratio of the kinetic energy to the interaction energy is
$\Ekin (t)/\Eint (t) \simeq b(t) - 1$ which is much larger than unity
for $t \gg 1/\omega$.
It follows from this analysis that for times typically larger than $1/\omega$, the expanding BEC
is a coherent superposition of almost non-interacting plane waves of momentum $k$:
%+++++++++++++++++++++++++++++++++++++++++%
\begin{equation}
\psi(z,t) = \int \frac{dk}{\sqrt{2\pi}} \widehat{\psi} (k,t) \textrm{e}^{i kz},
\label{eq:TFwf}
\end{equation}
%+++++++++++++++++++++++++++++++++++++++++%
the momentum distribution $\mathcal{D}(k)=|\hat{\psi}(k,t)|^2$ being stationary and
determined by the interaction-driven first expansion stage \cite{lsp2007}.

\paragraph{Anderson localization of quantum single particles in a correlated disordered potential~- }
Therefore, the interaction of each $k$-wave with the disordered potential can be treated independently.
According to the Anderson theory \cite{anderson1958}, the $k$-waves
will exponentially localize as a result of multiple scattering from the modulations of
the disordered potential.
In other words, each component $\textrm{e}^{i kz}$ in Eq.~(\ref{eq:TFwf}) will become a localized
function $\phi_k (z)$, characterized by an exponential decay at large distances:
$\ln|\phi_k(z)|\simeq -\gamma (k) |z|$,
where $\gamma (k)=1/\Lloc (k)$ is the so-called Lyapunov exponent, and $\Lloc (k)$ is
the localization length.
The Lyapunov exponent can be calculated analytically in a correlated disordered
potential using the phase formalism approach \cite{lifshits1988} 
(see also \ref{phaseformalism}).
At the lowest order of the Born expansion, which is valid provided that
$\gamma (k)\ll k$, \ie for
%+++++++++++++++++++++++++++++++++++++++++%
\begin{equation}
\Vr \sigmar \ll \frac{\hbar^2 k}{m} (k\sigmar)^{1/2},
\label{eq:lyapunovcond}
\end{equation}
%+++++++++++++++++++++++++++++++++++++++++%
we find the Lyapunov exponent
%+++++++++++++++++++++++++++++++++++++++++%
\begin{equation}
\gamma (k) \simeq 
\frac{\sqrt{2\pi}}{8\sigmar} 
\left(\frac{\Vr}{E}\right)^2
(k\sigmar)^2
\widehat{c} (2k\sigmar),
\label{eq:lyapunov1}
\end{equation}
%+++++++++++++++++++++++++++++++++++++++++%
where $E=\hbar^2k^2/2m$, and $\hat{c}(\kappa) = \int \frac{du}{\sqrt{2\pi}} c(u) \textrm{e}^{i\kappa u}$ is
the Fourier transform of the reduced correlation function $c(u)$ (see~\ref{potentials}).
A plot of the Lyapunov exponents versus the momentum $k$ is shown in Fig.~\ref{distgamma}b
for a speckle potential and for a Gaussian impurity model.

Deviations from a pure exponential decay of $\phi_k$ turn out to be important
here. They can be obtained using diagrammatic methods \cite{gogolin1976a,gogolin1976b}, and
one finds an integral formula for the average density of each $k$-wave:
%+++++++++++++++++++++++++++++++++++++++++%
\begin{eqnarray}
\langle| \phi_k (z) |^2\rangle 
& = & \frac{\pi^2 \gamma (k)}{2} \int_0^\infty du\ 
u\ \textrm{sinh} (\pi u) \left( \frac{1+u^2}{1+\textrm{cosh} (\pi u)} \right)^2
\nonumber \\
&&
\times  \textrm{exp} \{- 2 (1+u^2) \gamma (k) |z| \},
\label{eq:gogolin1}
\end{eqnarray}
%+++++++++++++++++++++++++++++++++++++++++%
where $\gamma(k)$ is given by Eq.~(\ref{eq:lyapunov1}).
Notice that at large distances ($|z| \gg 1/\gamma (k)$), Eq.~(\ref{eq:gogolin1})
reduces to
%+++++++++++++++++++++++++++++++++++++++++%
\begin{equation}
\langle| \phi_k (z) |^2 \rangle
\simeq  \left(\frac{\pi^{7/2}}{64\sqrt{2\gamma (k)}}\right) 
\times \frac{\exp\{-2\gamma (k)|z|\}}{|z|^{3/2}},
\label{eq:gogolin2}
\end{equation}
%+++++++++++++++++++++++++++++++++++++++++%
so that the exponential decay of the density of the localized single-particle states
is corrected by an algebraic decay $1/|z|^{3/2}$.

%-----------------------------------------%
%-----------------------------------------%

\paragraph{Anderson localization of the Bose-Einstein condensate~- }
In the regime of weak disorder defined by condition~(\ref{eq:lyapunovcond}), the Anderson localization
transforms each plane wave $\textrm{e}^{ikz}$ which appears in the superposition~(\ref{eq:TFwf}) into
the localized wave $\phi_z (z)=r(z)\sin[\theta(z)]$ where $\theta (z) \simeq kz$ 
and $r(z)$ is a slowly decaying envelope (see \ref{phaseformalism}).
Therefore, the Fourier transform of $\phi_k (z)$ is peaked around
$k$. It follows that the interaction of the $k$-wave with the disordered potential only weakly
affects the momentum distribution of the BEC. 
Hence, once each independent $k$-wave is localized, the density of the BEC is given by the equation
%+++++++++++++++++++++++++++++++++++++++++%
\begin{equation}
n_0(z) \simeq \left\langle \left| \int \frac{dk}{\sqrt{2\pi}} \hat{\psi}(k,t) \phi_k (z) \right|^2 \right\rangle
\label{eq:locBEC0}
\end{equation}
%+++++++++++++++++++++++++++++++++++++++++%
with $|\hat{\psi}(k,t)|^2 \simeq \mathcal{D}(k)$.
Assuming that the phases of the functions $\phi_k (z)$, which are determined by
the local properties of the disordered potential and by the evolution time,
are random, uncorrelated functions for different momenta, \ie 
$\langle \phi_{k'}^* (z) \phi_{k} (z) \rangle \simeq 2\pi \delta (k-k')$ , the density
of the BEC reduces to
%+++++++++++++++++++++++++++++++++++++++++%
\begin{equation}
n_0(z) \simeq
2\int_0^{\infty} dk \mathcal{D}(k) \langle | \phi_k (z) |^2\rangle
\label{eq:locBEC}
\end{equation}
%+++++++++++++++++++++++++++++++++++++++++%
where we have used the properties
$\mathcal{D} (k)=\mathcal{D} (-k)$ and $\langle | \phi_k (z) |^2\rangle = \langle | \phi_{-k} (z) |^2\rangle$.
This formula is transparent and contains the main ingredients of the Anderson localization of an interacting
BEC in a disordered potential. In a first stage, the interactions drive the expansion of the BEC and determine
the momentum distribution $\mathcal{D}(k)$. In a second stage, the interactions vanish and the BEC is formed of
a superposition of plane waves of energies $E=\hbar^2k^2/2m$. Then, each $k$-wave localizes with its own localization
length $\Lloc(k)=1/\gamma(k)$.
We will show in the next paragraphs that the precise localization properties of the BEC which are determined
by the integral~(\ref{eq:locBEC}), strongly depend on the correlation function of the disordered potential.
This is reminiscent of the strong dependence of the single-particle Lyapunov exponent $\gamma(k)$ on the
correlation function.

	%%%%%%%%%%%%%%%%%%%%%%%%%%%%%%%%%%%
	\subsection{Anderson localization of an expanding Bose-Einstein condensate in an impurity model of disorder}
\label{anderson:impurity}
Let us consider the case of the impurity model of disorder described in the \ref{potentials:impurity}.
It is made of a series of Gaussian peaks of width $w$ and amplitude $V_0$, all identical and spread
randomly along the $z$ axis. This potential is generic and in the limit
$w \rightarrow 0$ with $V_0w$ fixed, we recover the widely used $\delta$-correlated (\ie un-correlated) disorder made
of a random series of $\delta$ peaks used in a number of theoretical investigation of disordered
systems \cite{lifshits1988}. This Gaussian impurity model of disorder can also be implemented using the
so-called impurity atom technique with ultracold atomic gases \cite{gavish2005,paredes2005}. In this case,
the Fourier transform of the reduced correlation function reads
%+++++++++++++++++++++++++++++++++++++++++%
\begin{equation}
\hat{c}(\kappa) = \sqrt{\pi/2} \exp (-\kappa^2/4),
\label{correlTFimpurity}
\end{equation}
%+++++++++++++++++++++++++++++++++++++++++%
and the amplitude and correlation length are 
$\Vr=\sqrt{\frac{w}{d}}V_0$ 
and $\sigmar=2w$
respectively (see \ref{potentials:impurity}).
Inserting Eq.~(\ref{correlTFimpurity}) into Eq.~(\ref{eq:lyapunov1}), we find
%+++++++++++++++++++++++++++++++++++++++++%
\begin{equation}
\gamma(k) = \frac{\pi m^2\Vr^2\sigmar}{2\hbar^4 k^2} \exp[-(k\sigmar)^2]
\label{eq:lyapunovimpurity}
\end{equation}
%+++++++++++++++++++++++++++++++++++++++++%
which is plotted in Fig.~\ref{distgamma}b (blue, dotted line).

Using Eqs.~(\ref{momentthstat}),(\ref{eq:gogolin1}),(\ref{eq:locBEC}),(\ref{eq:lyapunovimpurity}),
we now calculate the density profile of the localized BEC.
Since the density profile $n_0(z)$ is the sum over $k$ of the functions $\langle | \phi_k (z) |^2 \rangle$
which decay exponentially with a rate $2\gamma (k)$, the long-tail behavior of $n_0(z)$ 
is mainly determined by the $k$-components with the smallest $\gamma (k)$, \ie those with $k$ close to 
the high-momentum cut-off $\kc=1/\xiini$.
Therefore, integrating in Eq.~(\ref{eq:locBEC}) we limit ourselves to the leading order terms
in Taylor series for $\mathcal{D} (k)$ and $\gamma (k)$ at $k$ close to $\kc$.
We find
%%%%%%%%%%%%%%%%%%%%%%%%%%%%%%%%%%%%%%%%%%%%%%%%%%%%%%%%%%%%%%%%%%%%%%%%%%%%%
\begin{eqnarray}
& & n_0(z) \propto \frac{\exp\{-2\gammaeff|z|\}}{|z|^{7/2}}
\label{eq:asymptimpurity1} \\
& \textrm{where~~} & \gammaeff = \gamma (k=1/\xiini).
\label{eq:asymptimpurity2}
\end{eqnarray}
%%%%%%%%%%%%%%%%%%%%%%%%%%%%%%%%%%%%%%%%%%%%%%%%%%%%%%%%%%%%%%%%%%%%%%%%%%%%

This means that the Anderson localization of an expanding, interacting BEC occurs,
provided that the disordered potential is
weak enough. In the case of a Gaussian impurity model of disorder, the density profile shows
an exponential decay with the effective Lyapunov exponent equal to the one of a single particle
of momentum $k=\kc=1/\xiini$ [see Eq.~(\ref{eq:asymptimpurity2})], \ie
%+++++++++++++++++++++++++++++++++++++++++%
\begin{equation}
\gammaeff = \frac{\pi/32}{\xiini} \left(\frac{\Vr}{\mu}\right)^2 
(\sigmar/\xiini) \exp[-(\sigmar/\xiini)^2].
\label{eq:lyapunovBECimpurity}
\end{equation}
%+++++++++++++++++++++++++++++++++++++++++%
This is a clear characteristics of Anderson localization of the BEC which can be observed in
experiments on ultracold atoms using direct imaging techniques.

	%%%%%%%%%%%%%%%%%%%%%%%%%%%%%%%%%%%
	\subsection{Anderson localization of an expanding Bose-Einstein condensate in a speckle potential}
\label{anderson:speckle}
Let us examine now the case of a speckle potential (see \ref{potentials:speckle}) for
which, in 1D, the Fourier transform of the reduced correlation~(\ref{corr:speckle}) function reads
%+++++++++++++++++++++++++++++++++++++++++%
\begin{equation}
\hat{c}(\kappa) = \sqrt{\pi/2} (1-\kappa/2) \Theta (1-\kappa/2),
\label{correlTFspeckle}
\end{equation}
%+++++++++++++++++++++++++++++++++++++++++%
where $\Theta$ is the Heaviside step function.
The case of a speckle potential is particularly interesting for two reasons.
First, it corresponds to the model
of disorder used in almost all present experiments with disordered BECs 
\cite{lye2005,clement2005,fort2005,clement2006,clement2007,chen2007}. Second,
we will see that speckle potentials offer much richer situations than the impurity model
discussed above due to peculiar long-range correlations \cite{lsp2007}.

One important feature of the speckle potential is the fact that the Fourier
transform~(\ref{correlTFspeckle}) of the correlation function has a finite support. After Eq.~(\ref{eq:lyapunov1}),
it results that the Lyapunov exponent vanishes for $k>1/\sigmar$, \ie that strong Anderson localization
occurs for non-interacting $k$-waves only for $k<1/\sigmar$ \cite{lsp2007}.
In other words, there is a 1D mobility edge at $1/\sigmar$ in the Born approximation.
Strictly speaking, higher orders in the Born expansion may provide a non-vanishing
Lyapunov exponent for $k>1/\sigmar$.
However, we have shown using direct numerical calculations that the localization length (inverse Lyapunov exponent)
for $k>1/\sigmar$ strongly exceeds typical sizes of ultracold atomic samples, so that we can
consider $k=1/\sigmar$ as an {\it effective mobility edge} in our problem \cite{lsp2007}.

It follows that a part of the expanding BEC (\ie its Fourier components with $k>1/\sigmar$) expands to infinity
while all the Fourier components with $k<1/\sigmar$ localize exponentially with the $k$-dependent
Lyapunov exponent
%+++++++++++++++++++++++++++++++++++++++++%
\begin{equation}
\gamma(k) = \frac{\pi m^2\Vr^2\sigmar}{2\hbar^4 k^2} (1 - k\sigmar)\Theta(1 - k\sigmar),
\label{eq:lyapunovspeckle}
\end{equation}
%+++++++++++++++++++++++++++++++++++++++++%
found by inserting Eq.~(\ref{correlTFspeckle}) into Eq.~(\ref{eq:lyapunov1}).
Equation~(\ref{eq:lyapunovspeckle}) is plotted in Fig.~\ref{distgamma}b (solid, red line)
and show that for a 1D speckle potential, the high-momentum cut-off
$\kc=\min\{1/\xiini,1/\sigmar\}$ in the integral formula~(\ref{eq:locBEC}) for the BEC density
is twofold. The cut-off $k=1/\xiini$ is related to the momentum distribution of the expanding BEC
and is due to the initial atom-atom interactions,
while the cut-off $k=1/\sigmar$ is related to the correlation function of the 1D speckle potential
and is due to the peculiar finite range correlations of the disordered potential.
Now, two very different situations must be distinguished \cite{lsp2007}.

For $\xiini>\sigmar$, the high-momentum cut-off $\kc$ is provided by the momentum distribution.
In this case all non-interacting functions $\langle | \phi_k (z) |^2 \rangle$ are exponentially localized
with a finite Lyapunov exponent, $\gamma (k) \geq \gamma (1/\xiini) > 0$.
This situation is then similar to the case of the Gaussian impurity model and,
integrating Eq.~(\ref{eq:locBEC}), we find
%%%%%%%%%%%%%%%%%%%%%%%%%%%%%%%%%%%%%%%%%%%%%%%%%%%%%%%%%%%%%%%%%%%%%%%%%%%%%
\begin{eqnarray}
& & n_0(z) \propto \frac{\exp\{-2\gammaeff|z|\}}{|z|^{7/2}}
\label{eq:asympt1speckle1} \\
& \textrm{where~~} & \gammaeff = \gamma (k=1/\xiini).
\label{eq:asympt1speckle2}
\end{eqnarray}
%%%%%%%%%%%%%%%%%%%%%%%%%%%%%%%%%%%%%%%%%%%%%%%%%%%%%%%%%%%%%%%%%%%%%%%%%%%%
Finally, the BEC density profile is exponentially localized with the effective Lyapunov exponent
%+++++++++++++++++++++++++++++++++++++++++%
\begin{equation}
\gammaeff = \frac{\pi/32}{\xiini} \left(\frac{\Vr}{\mu}\right)^2
(\sigmar/\xiini)(1-\sigmar/\xiini) \Theta (1-\sigmar/\xiini).
\label{eq:lyapunovBECspeckle}
\end{equation}
%+++++++++++++++++++++++++++++++++++++++++%

For $\xiini<\sigmar$, the situation is completely different. In this case, the cut-off $\kc$ is
provided by the correlation function, and since $\gamma (k=1/\sigmar)=0$, the relevant Lyapunov exponents
($\gamma (k)$ for all $k<1/\sigmar$) do not have a finite lower bound.
Then, integrating Eq.~(\ref{eq:locBEC}), we find that the BEC density profile is not exponentially localized
but rather shows an {\it algebraic} decay \cite{lsp2007}:
%+++++++++++++++++++++++++++++++++++++++++%
\begin{equation}
n_0(z) \propto \frac{1}{|z|^{2}}.
\label{eq:asympt1speckle}
\end{equation}
%+++++++++++++++++++++++++++++++++++++++++%

\begin{figure}[t!]
\begin{center}
\includegraphics[width=8.5cm]{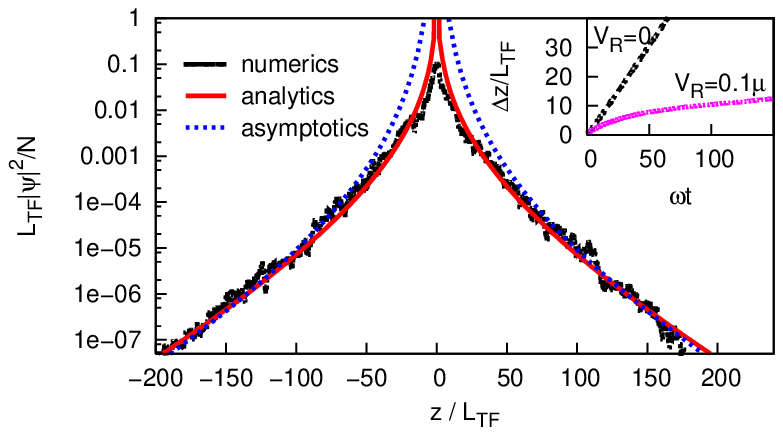}
\hspace{0.5cm}\includegraphics[width=10.cm]{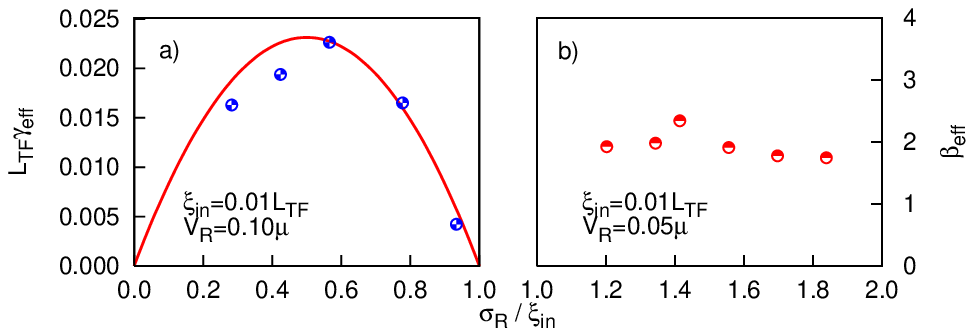}
\end{center}
\caption{(color online)
Upper panel: Density profile of the localized BEC in a
speckle potential at $t=150/\omega$. Shown are the numerical data (black points),
the fit of the result from Eqs.~(\ref{momentthstat}), (\ref{eq:gogolin1}) 
and (\ref{eq:locBEC}) [red solid line], and the fit of
the asymptotic formulas~(\ref{eq:asympt1speckle1}),(\ref{eq:asympt1speckle2}) [blue dotted line].
Inset: Time evolution of the rms size of the BEC.
The parameters are $\Vr=0.1\mu$, $\xiini=0.01 \LTF$,
and $\sigmar=0.78\xiini$.
Lower panel: a) Lyapunov exponent $\gammaeff$ in units of $1/\LTF$ 
for the localized BEC in a speckle potential, in the regime $\xiini>\sigmar$.
The solid line is $\gamma(1/\xiini)$ from Eq.~(\ref{eq:lyapunovBECspeckle}).
b) Exponent of the power-law decay of the localized BEC in the regime 
$\xiini<\sigmar$. The parameters are indicated in the figure.}
\label{ALspeckle}
\end{figure}

We now present numerical results performed within the Gross-Pitaevskii approximation for the
expansion of a BEC in a speckle potential \cite{lsp2007}. The inset of Fig.~\ref{ALspeckle}
(upper panel) shows that the expansion is strongly suppressed and for long times, the BEC
density profile is localized as shown in Fig.~\ref{ALspeckle} (upper panel). Let us
discuss now the behavior of the tails.

%-----------------------------------------%
%-----------------------------------------%

For $\xiini>\sigmar$, the density profile obtained numerically is found to be exponentially localized.
In addition, fitting the function $n_0(z) \propto \exp\{-2\gammaeff|z|\}/|z|^{7/2}$ to the numerical results
with the amplitude and $\gammaeff$ as fitting parameters, we find that the results for $\gammaeff$ are in excellent
agreement with the prediction~(\ref{eq:lyapunovBECspeckle}) as shown in Fig.~\ref{ALspeckle}a (lower panel).

For $\xiini<\sigmar$, we find that the density profile decays algebraically. We fit the function
$n_0 (z) \propto 1/|z|^{\beta_{\tiny eff}}$ with the amplitude and $\beta_{\tiny eff}$ as fitting parameters,
and we find that $\beta_{\tiny eff} \simeq 2$ in agreement with the prediction~(\ref{eq:asympt1speckle})
as shown in Fig.~\ref{ALspeckle}b (lower panel).

%%%%%%%%%%%%%%%%%%%%%%%%%%%%%%%%%%%%%%%%%%%%%%%%%%%%%%%%%%%%%%%%%%%%%%
\section{Conclusion and perspectives}
\label{conclusion}

In summary, we have theoretically investigated the localization of an expanding
1D BEC with repulsive atom-atom interactions characterized by the initial healing length
$\xiini$ in a disordered potential of finite correlation length $\sigmar$.
We have restricted our study to the regime where the initial interactions of the trapped BEC
dominate over the kinetic energy and the disorder, a situation relevant to almost
all current experiments with disordered BECs
\cite{lye2005,clement2005,fort2005,schulte2005,clement2006,clement2007,chen2007}.
When the BEC is released from the trapping potential while keeping the disordered potential on,
we find a strong suppression of the expansion, similar to earlier experimental observations
\cite{clement2005,fort2005,clement2006}.
We have shown that this localization effect has completely different causes depending on
the strength of the disorder.

\paragraph{Strong disorder -}
The case of strong disorder corresponds to the situation where several modulations of the disordered
potential individually are strong-enough to induce an almost total reflection of
noninteracting particles of energy equal to the typical expansion
energy per particle of the BEC.
In particular this case is relevant to the experiments of 
Refs.~\cite{clement2005,fort2005,clement2006} and to the numerics of 
Refs.~\cite{clement2005,modugno2005,akkermans2006}.
In this case, the localization results from a {\it disorder-induced trapping} \cite{clement2005},
whose scenario involves two processes: 
(i) the fragmentation of the core of the BEC on one hand, and
(ii) classical total reflections from single large modulations of the disordered potential
on the other hand.
In the core of the BEC, the interactions remain important during the initial expansion stage.
The BEC is in a quasi-static Thomas-Fermi regime with an effective chemical potential which
first slowly decreases during the expansion. When the BEC fragments, the expansion of the
core of the BEC stops.
In the tails of the BEC, the interactions are negligible and the particles undergo
multiple scattering from the modulations of the disordered potential but the expansion
is ultimately stopped by single large modulations.
Hence, in the case of strong disorder, the localization is {\it not} related to Anderson localization.

\paragraph{Weak disorder -}
In the case of weak-enough disorder, the probability of modulations of the disordered potential
such that the reflection of a particle of typical energy $\mu$ approaches unity is negligible,
and Anderson localization can show up in an expanding BEC. The scenario is then as follows \cite{lsp2007}.
In a first stage, the interaction energy dominates over both the kinetic energy and the disorder,
and drives the initial expansion of the BEC. After a typical time of $1/\omega$ where
$\omega$ is the frequency of the initial trapping potential, the interaction energy vanishes
and the momentum distribution of the expanding BEC becomes stationary.
At this stage, the interactions can be neglected and the BEC wavefunction is a superposition
of (almost) non-interacting waves of momentum $k$. Each $k$-wave Anderson localizes with its own
$k$-dependent localization length $\Lloc (k)$.
The density profile of the localized BEC is thus the
superposition of the localized $k$-waves and strongly depends on the correlation function of the
disordered potential.

The case of speckle potentials is particularly interesting as the Fourier transform of their
correlation function has a finite support. It follows that the localization of the expanding BEC
is exponential only for $\xiini>\sigmar$ (in the lowest order of the Born expansion,
see Sec.~\ref{anderson}). In the opposite situation ($\xiini<\sigmar$), the density
profile of the BEC decays algebraically as $1/|z|^2$.
Therefore, for speckle potentials, there is an {\it effective mobility edge} at $\xiini=\sigmar$
for the Anderson localization of an expanding, interacting BEC \cite{lsp2007}.

\paragraph{Perspectives -}
Our results suggest that the 1D Anderson localization can be observed in an interacting BEC
(initially in the Thomas-Fermi regime) expanding in a disordered potential in experiments similar
to those reported in Refs.~\cite{clement2005,fort2005,clement2006}.
We stress that special attention should be paid to using weak-enough disorder to allow
the dilution of the BEC during the first expansion stage and to avoid strong reflections from
large modulations of the disordered potential.
In addition, we have shown that the correlation function of the disordered potential
plays a crucial role for the localization properties of the BEC.

However, a couple of challenges have to be taken up to observe Anderson localization in
current experiments with expanding BECs. Indeed, in addition to being able to produce
long-enough expansions and to measure very small densities, it appears that both
disorder and interactions have to be carefully controlled.

In this respect, using disordered potentials created by optical speckle patterns is
particularly promissing. On one hand, from a practical point of view, the correlation
functions of speckle potentials are very well controlled and can be designed almost at will.
This allows for a direct comparison between experimental observations and theoretical predictions.
On the other hand, for a 1D speckle potential, the Fourier transform of the correlation
function has a cut-off at $k=1/2\sigmar$ which induces an {\it effective mobility edge} at
$k=1/\sigmar$ for single-particles and, correspondingly, at $\xiini = \sigmar$ for an
expanding, interacting BEC. The presence of this effective mobility edge gives rise to
two qualitatively different regimes, which might be observed in experiments, namely
{\it exponential localization} for $\xiini>\sigmar$ and
{\it algebraic decay} of the BEC density profile for $\xiini<\sigmar$.

Equation~(\ref{eq:lyapunovBECspeckle}) shows that for a speckle potential, the stronger
localization is obtained for $\xiini=3\sigmar/2$ and for this value, $\Lloc$ increases
with $\sigmar$. It is thus more favourable to work with the shortest correlation
length of the disordered potential. To date, correlation lengths about
$\sigmar \simeq 0.3\mu$m have been produced experimentally \cite{clement2006}.

The first experiments on the expansion of a BEC in a speckle potential \cite{clement2005,fort2005,clement2006}
were operated at $\xiini \sim \sigmar/10$.
In order to reach the regime of Anderson localization, it is crucial to lower the
interaction energy of the initial condensate in order
to increase the healing length $\xiini$. One can either use a Feshbach resonance
to directly control the atom-atom interaction strength or lower the density
by lowering the number of atoms and/or the radial confinement of the magnetic guide used in
Refs.~\cite{clement2005,fort2005,clement2006}. In the latter case, the dynamics
of the expanding BEC in the radial direction can play a role.
As the radial confinement is kept
during the expansion of the BEC, we expect the radial dynamics to be slow compared
to the longitudinal expansion. As a result, the radial profile of the BEC
would follow adiabatically the local 1D (longitudinal) density, reducing the
dynamics to a quasi-1D problem. However, a part of the potential and interaction
energies associated to the smooth radial confinement
will be converted into longitudinal kinetic energy during the expansion.
If the BEC is initially in the 3D Thomas-Fermi regime ($\mu \gg \hbar\omega_\perp$),
the effective de Broglie wavelength of the 1D expansion can be expected to be
$\ldB \simeq 0.85 \xiini$ to be compared to $\xiini$ in the pure
1D case studied here.
Hence, for elongated BECs, we do not expect strong differences compared to the 1D situation
we have studied in the present work. A more detailed discussion would require further
investigation which is beyond the scope of this paper.

We expect that this work will pave the way for the experimental observation of those
non-trivial 1D localization properties. This work could also be extended to higher dimensions
(namely 2D and 3D geometries) \cite{shapiro2007,skipetov2008}
where similar scenarios can be expected but where localization
properties would be significantly different.

%%%%%%%%%%%%%%%%%%%%%%%%%%%%%%%%%%%%%%%%%%%%%%%%%%%%%%%%%%%%%%%%%%%%%%
\ack
We are indebted to P.~Chavel, D.~Gangardt, G.V.~Shlyapnikov, M.~Lewenstein,
J.A.~Retter and A.~Varon for fruitful discussions.
This work was supported by
the Centre National de la Recherche Scientifique (CNRS),
the D\'{e}l\'{e}\-gation G\'{e}n\'{e}rale de l'Armement (DGA),
the Minist\`ere de l'Enseignement National, de la Recherche et de la Technologie (MENRT),
the Agence Nationale de la Recherche (ANR, contract NTOR-4-42586),
and the programme QUDEDIS of the European Science Foundation (ESF).
The Atom Optics group at LCFIO is a member of the Institut Francilien 
de Recherche sur les Atomes Froids (IFRAF).

%%%%%%%%%%%%%%%%%%%%%%%%%%%%%%%%%%%%%%%%%%%%%%%%%%%%%%%%%%%%%%%%%%%%%%
\appendix

%%%%%%%%%%%%%%%%%%%%%%%%%%%%%%%%%%%%%%%%%%%%%%%%%%%%%%%%%%%%%%%%%%%%%%
\section{Random and periodic potentials}
\label{potentials}
\begin{figure}[t!]
\begin{center}
\infig{40em}{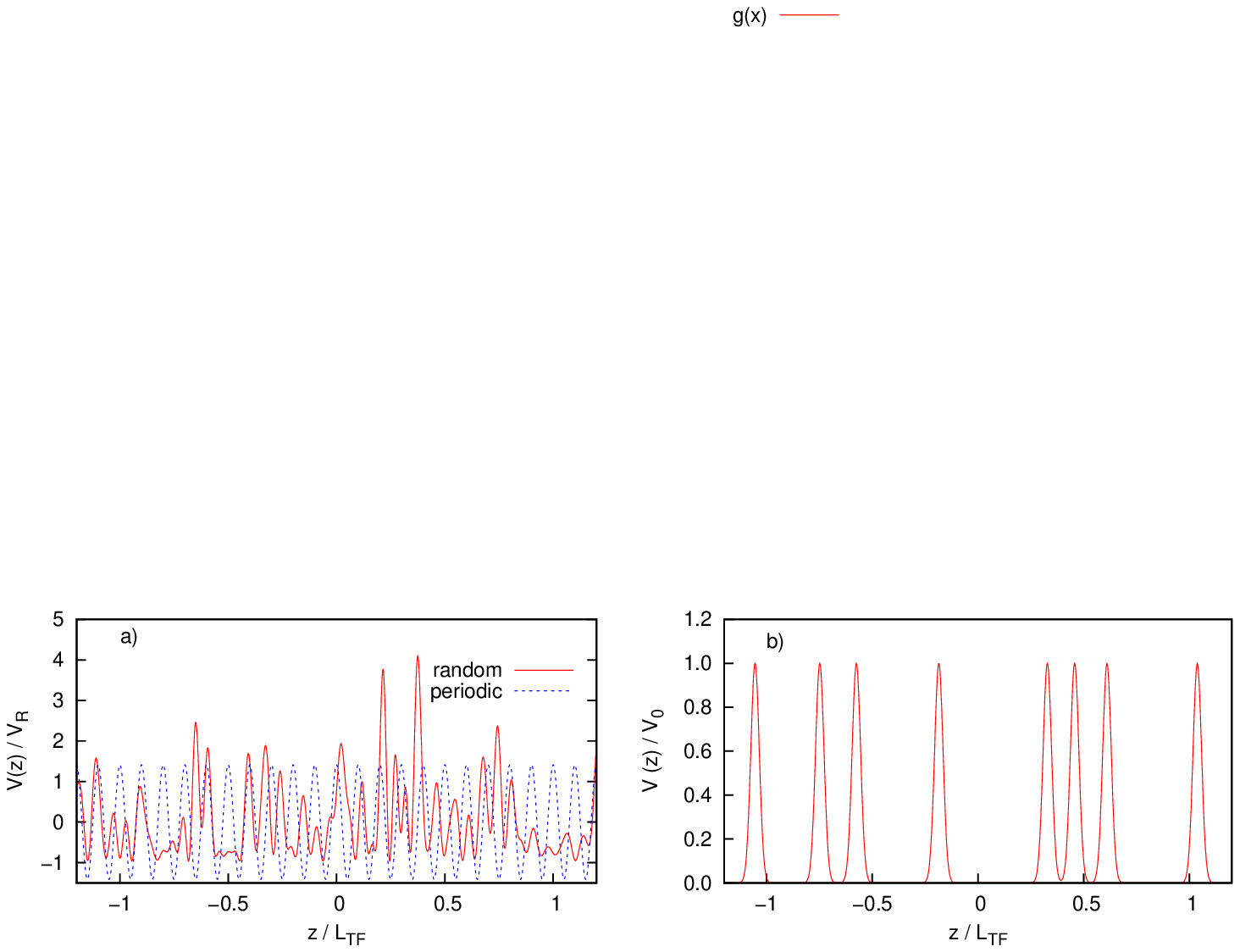}
\end{center}
\caption{(color online)
Typical realizations of the potentials $\Vopt (z)$ considered in this work.
a) Speckle and periodic potentials:
the solid line (red online) shows a typical realization of a disordered speckle potential with
$\sigmar \simeq 1.39\times 10^{-2} \LTF$ and the dotted line (blue online) shows the periodic potential
with $\lambda = 10^{-1} \LTF$.
b) Gaussian impurity model of disorder with $w=2\times 10^{-2} \LTF$.
} 
\label{potential}
\end{figure}

In this appendix, we give a couple of details about the inhomogeneous potentials $\Vopt (z)$ to which the
atoms are subjected (see section~\ref{equilibrium:system}) \ie
a disordered speckle potential,
a Gaussian impurity model of disorder and
a periodic potential.
We plot in Fig.~\ref{potential} typical realizations of these potentials.

Generally, we write the potential $\Vopt (z)$ as
%+++++++++++++++++++++++++++++++++++++++++%
\begin{equation}
\Vopt (z) = \Vr v (z/\sigmar)
\label{eq:scaledpot}
\end{equation}
%+++++++++++++++++++++++++++++++++++++++++%
and the spatial auto-correlation function
$C(z)=\langle \Vopt (z'+z) \Vopt (z') \rangle - \langle \Vopt \rangle^2$
as
%+++++++++++++++++++++++++++++++++++++++++%
\begin{equation}
C (z) = \Vr^2 c (z/\sigmar)
\label{eq:scaledcorr}
\end{equation}
%+++++++++++++++++++++++++++++++++++++++++%
where $\Vr$ is the typical amplitude of potential,
$\sigmar$ is the typical space scale (correlation length in the case of disordered potentials)
and $v (u)$ is a given function characteristic of the model of inhomogeneous potential.
In this work, we assume that $\langle v (z) \rangle = 0$ where
$\langle . \rangle$ represents averaging over realizations of disordered potentials or
spatial averaging.

%-----------------------------------------%
%-----------------------------------------%

	%%%%%%%%%%%%%%%%%%%%%%%%%%%%%%%%%%%
	\subsection{Speckle disordered potential}
\label{potentials:speckle}
The main model of disorder we consider is the speckle potential, as it is the one
used in many experimental studies of disordered Bose-Einstein condensates
\cite{lye2005,clement2005,fort2005,clement2006,clement2007,chen2007}.
In brief a speckle pattern is formed by diffraction of a laser beam through a rough
plate. The intensity of the speckle pattern is proportional to the intensity of the incident laser and the
correlation function is determined by the transmission of the diffusive plate
\cite{goodman1975,goodman2007} (see Ref.~\cite{clement2006}
for practical realizations in the context of disordered BECs).
A disordered potential (\eg a speckle potential) is characterized by its statistical
properties, mainly the single-point intensity distribution and the two-point correlation
function.

Within the scaling defined above [see Eqs.~(\ref{eq:scaledpot}),(\ref{eq:scaledcorr})],
a speckle potential is represented by a random function $v(u)$ whose single-point statistical distribution
is a decaying exponential
%+++++++++++++++++++++++++++++++++++++++++%
\begin{eqnarray}
& & \mathcal{P}[v(u)] = \exp[-\{v(u)+1\}]
\textrm{~~~~for~} v(u) \ge -1 \label{probaintens} \\
\textrm{and~~} & & \mathcal{P}[v(u)] = 0 \nonumber
\textrm{~~~~otherwise}.
\end{eqnarray}
%+++++++++++++++++++++++++++++++++++++++++%
The laser intensity pattern creates an inhomogeneous light shift for the atoms (see Fig.~\ref{potential}a).
For a laser which is blue-detuned compared to the atomic resonance line, we have
$\Vr>0$ and for red detuning $\Vr<0$ \cite{grimm2000}.

For the numerical calculations presented in the paper, we numerically generate
a 1D speckle pattern using a method similar to the one described in Refs.~\cite{horak1998,huntley1989}
in 1D and corresponding to the following reduced correlation function:
%+++++++++++++++++++++++++++++++++++++++++%
\begin{equation}
c (u) = \sinc (u)^2.
\label{corr:speckle}
\end{equation}
%+++++++++++++++++++++++++++++++++++++++++%

\begin{figure}[t!]
\begin{center}
\includegraphics[width=8.5cm]{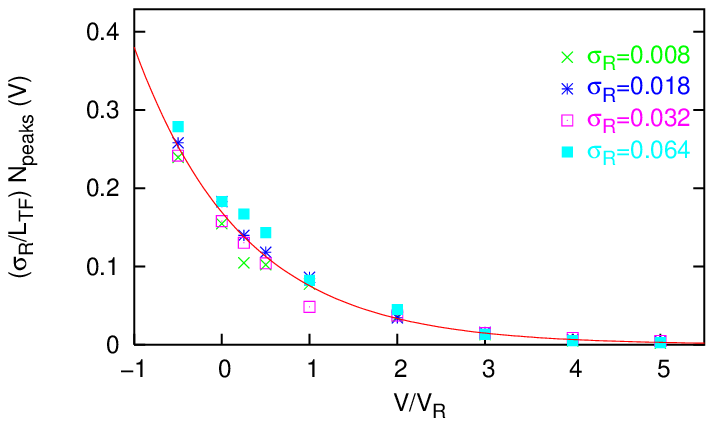}
\end{center}
\caption{(color online)
Average number of peaks within a range $\LTF$ with amplitude
larger than $V$ for different values of the correlation length of the
disordered potential and comparison to Eq.~(\ref{picseqtext}).
} 
\label{pics}
\end{figure}

Another useful characteristics of the speckle potential in the case of blue detuning
($\Vr>0$) is the average number of peaks
with an intensity larger than a given value $V$
within a given region of length $\LTF$ (see section~\ref{transport:scenario}).
Elaborate methods to compute a number of characteristics of speckle potentials can be
found in Refs.~\cite{goodman1975,goodman2007}.
Here, we use a simple approximation which suits our purpose.
From the probability distribution~(\ref{probaintens}), we easily find that the
probability density that the local disordered potential is larger than a given value $V$ is
$P(V) = \exp[-(V+\Vr)/\Vr]$. Now, the density of peaks
(local maxima of the disordered potential) is $1/d$ 
where $d \propto \sigmar$ is the typical distance between two peaks.
Therefore, typically, the number of peaks $\Npeaks$ whithin a region of length $\LTF$
with intensity larger than $V$ scales as
$\frac{\LTF}{d} \exp\left(-\frac{V+\Vr}{\Vr}\right)$.
However, this rough estimate does not take into account the interplay between
the local intensity distribution and the finite correlation length of the disordered potential.
In the simulated speckle potentials, we find that the typical number of peaks
with intensity larger than $V$ in a $\LTF$-long region can be approximated by
%+++++++++++++++++++++++++++++++++++++++++%
\begin{equation}
\Npeaks (V) \simeq \alpha \left(\frac{\LTF}{\sigmar}\right) \exp\left[ -\beta \frac{V+\Vr}{\Vr} \right]
\label{picseqtext}
\end{equation}
%+++++++++++++++++++++++++++++++++++++++++%
with $\alpha \simeq 0.30$ and $\beta \simeq 0.75$,
with very good accuracy as shown in Fig.~\ref{pics}.

%-----------------------------------------%
%-----------------------------------------%

	%%%%%%%%%%%%%%%%%%%%%%%%%%%%%%%%%%%
	\subsection{Impurity model of disorder}
\label{potentials:impurity}
A model of disorder which is very popular in theoretical studies of quantum disordered
system is the impurity model \cite{lifshits1988}:
%+++++++++++++++++++++++++++++++++++++++++%
\begin{equation}
V(z) = V_0 \sum_j g(z-Z_j),
\label{toypot}
\end{equation}
%+++++++++++++++++++++++++++++++++++++++++%
where $g$ is a real-valued function peaked at $z=0$, of width $w$ and such that $0 \leq g(z) \leq 1$.
The locations of the impurities $Z_j$ are random and their average distance is denoted $d$.
The disordered potential is then formed of a series of impurities, all identical but
randomly displaced along the $z$ axis (see Fig.~\ref{potential}b).
Here, we consider Gaussian-shaped impurities:
%+++++++++++++++++++++++++++++++++++++++++%
\begin{equation}
g(z) = \exp ( -z^2/2w^2 ).
\label{toypotgauss}
\end{equation}
%+++++++++++++++++++++++++++++++++++++++++%
This potential can be realized using ultracold atoms (of another species than the
BEC) trapped in the Wannier states of the fundamental Bloch band of an optical lattice
\cite{gavish2005,paredes2005,massignan2006}.

From Eq.~(\ref{toypot}), we find that the statistical average of the potential~(\ref{toypot}) is
%+++++++++++++++++++++++++++++++++++++++++%
\begin{equation}
\langle V \rangle = \frac{V_0}{d} \int dx\ g(x) = \sqrt{2\pi} V_0 \left(\frac{w}{d}\right),
\label{toypotav}
\end{equation}
%+++++++++++++++++++++++++++++++++++++++++%
and the correlation function is
%+++++++++++++++++++++++++++++++++++++++++%
\begin{equation}
C(z) = \frac{V_0^2}{d} \int dx\ g(x) g(x+z) = \Vr^2 c(z/\sigmar)
\label{toycorr}
\end{equation}
%+++++++++++++++++++++++++++++++++++++++++%
with $c(u)=\sqrt{\pi} \exp(-x^2)$, $\Vr = \sqrt{\frac{w}{d}} V_0$ and $\sigmar=2w$.

	%%%%%%%%%%%%%%%%%%%%%%%%%%%%%%%%%%%
	\subsection{Periodic potential}
\label{potentials:periodic}
For periodic potentials (see Fig.~\ref{potential}a), we use
%+++++++++++++++++++++++++++++++++++++++++%
\begin{equation}
\Vopt(z) = \sqrt{2} \Vr \cos(2\pi z /\lambda)
\label{periopot}
\end{equation}
%+++++++++++++++++++++++++++++++++++++++++% 
corresponding to the mean value $\langle \Vopt \rangle = 0$ and the standard deviation $\Delta \Vopt = |\Vr|$.
Such potentials are currently realized using interference patterns of laser
beams in geometries with tunable lattice spacings \cite{lattices2000,fallani2005}.

%%%%%%%%%%%%%%%%%%%%%%%%%%%%%%%%%%%%%%%%%%%%%%%%%%%%%%%%%%%%%%%%%%%%%%
\section{Phase formalism for the calculation of Lyapunov exponents in 1D disordered potentials with finite-range correlations}
\label{phaseformalism}
In this section, we outline the phase formalism method used to calculate the Lyapunov exponent (inverse localization
length) of a non-interacting particle of energy $E=\hbar^2k^2/2m$ in a 1D disordered potential with finite-range
correlations \cite{lifshits1988}. The idea consists in calculating the propagation of the particle using a perturbation
on the {\it phase} of the wavefunction $\psi (z)$.
Notice that the perturbation is {\it not} on the wavefunction itself.

The starting point is the Schr\"odinger equation
%+++++++++++++++++++++++++++++++++++++++++%
\begin{equation}
E\psi (z) = \frac{-\hbar^2}{2m} \frac{d^2}{dz^2} \psi (z) + V(z)\psi (z).
\label{schrodinger}
\end{equation}
%+++++++++++++++++++++++++++++++++++++++++%
Without any loss of generality, we can write the wavefunction and its spatial derivative in the form
%+++++++++++++++++++++++++++++++++++++++++%
\begin{eqnarray}
\psi (z) = r(z)\sin[\theta (z)] \label{phaseF1} \\
\psi' (z) = k r(z)\cos[\theta (z)] \label{phaseF2}
\end{eqnarray}
%+++++++++++++++++++++++++++++++++++++++++%
where $r(z)$ and $\theta(z)$ represent the amplitude and the phase of $\psi (z)$, respectively.
Substituting Eqs.~(\ref{phaseF1}),(\ref{phaseF2}) into Eq.~(\ref{schrodinger}),
we find the coupled equations
%+++++++++++++++++++++++++++++++++++++++++%
\begin{eqnarray}
\theta' (z) = k - \frac{2m V(z)}{\hbar^2 k} \sin^2 [\theta (z)] \label{phaseEq1} \\
\ln [r(z)/r(0)] = \int_0^z dz'\ \frac{m V(z')}{\hbar^2 k} \sin [2\theta (z')] \label{phaseEq2}
\end{eqnarray}
%+++++++++++++++++++++++++++++++++++++++++%
Notice that the amplitude $r(z)$ is not involved in Eq.~(\ref{phaseEq1}). It follows that for weak disorder
[see condition~(\ref{phasecond})],
it can be solved easily in the lowest order of a perturbation series of the phase $\theta (z)$.
We write
$\theta (z) = \theta_0 + kz + \delta \theta (z)$ and we find
%+++++++++++++++++++++++++++++++++++++++++%
\begin{equation}
\theta (z) \simeq \theta_0 + kz - \int_0^z dz'\ \frac{2m V(z')}{\hbar^2 k} \sin^2 [\theta_0 + kz']
\label{phaseS1}
\end{equation}
%+++++++++++++++++++++++++++++++++++++++++%
in the Born approximation (lower order).
Finally, substituting Eq.~(\ref{phaseS1}) into Eq.~(\ref{phaseEq2}), we find, in the limit
$|z| \rightarrow \infty$
%+++++++++++++++++++++++++++++++++++++++++%
\begin{equation}
\ln [r(z)/r(0)] = + \gamma (k) |z|
\label{phaseS2}
\end{equation}
%+++++++++++++++++++++++++++++++++++++++++%
where $\gamma (k) = \frac{m}{4\hbar^2 E} \int_{-\infty}^{+\infty} dz\ C(z) \cos (2kz)$,
where $C(z)$ is the correlation function of the disordered potential [see Eq.~(\ref{eq:scaledcorr})]
or equivalently
%+++++++++++++++++++++++++++++++++++++++++%
\begin{equation}
\gamma (k) \simeq 
\frac{\sqrt{2\pi}}{8\sigmar} 
\left(\frac{\Vr}{E}\right)^2
(k\sigmar)^2
\widehat{c} (2k\sigmar),
\label{eq:lyapunov1bis}
\end{equation}
%+++++++++++++++++++++++++++++++++++++++++%
is the Lyapunov exponent.

Notice that the solution~(\ref{phaseS2}) corresponds to an exponential {\it increase} of the envelope
of the wavefunction $\psi (z)$. In general, the solution of the Schr\"odinger equation in a disordered potential
is the sum of an exponentially increasing function and of an exponentially decaying function. In a finite
system, boundary conditions fix the coefficients and one finds true localized states
(\ie wavefunctions with an exponentially decaying envelope for both $z \rightarrow +\infty$ and $z \rightarrow -\infty$).
In a time-dependent
propagation scheme, the conservation of probability also imposes that the coefficient of the exponentially
increasing function vanishes. In the present calculation, since we do not impose boundary conditions,
only the exponentially increasing function remains at infinite distance.

In spite of this unimportant limitation, this technique provides a very useful
analytic formula for the Lyapunov exponent, valid for any weak 1D disordered potential, possibly with
finite-range correlations.
An important point is that the phase formalism technique \cite{lifshits1988} clarifies the Anderson
localization effect in 1D disordered potentials.
Hence, on small distances (say of the order of $\sigmar$), the interaction of the wavefunction with
the disordered potential induces a small perturbation of the phase $\theta (z)$ [see Eq.~(\ref{phaseEq1})], but
hardly affect the amplitude $r(z)$. Nevertheless, the coupling between the phase and the amplitude
[see Eq.~(\ref{phaseEq2})] is crucial and induces at large distances (say of the order of $\Lloc=1/\gamma$)
an exponential envelope, characteristic of Anderson localization.
Finally, the non-interacting localized state of energy $E$ is essentially a plane wave of wavenumber
$k=\sqrt{2mE/\hbar^2}$ modulated by an exponential envelope. This makes clear the condition of application
of the phase formalism approach which requires $\gamma (k) \ll k$ in order for the phase to be only
weakly perturbed. It follows for Eq.~(\ref{eq:lyapunov1bis}) that this condition reduces to
%+++++++++++++++++++++++++++++++++++++++++%
\begin{equation}
\Vr \sigmar \ll \frac{\hbar^2 k}{m} (k\sigmar)^{1/2}
\label{phasecond}
\end{equation}
%+++++++++++++++++++++++++++++++++++++++++%
where $\Vr$ is the amplitude and $\sigmar$ is the correlation length of the disordered potential.

%%%%%%%%%%%%%%%%%%%%%%%%%%%%%%%%%%%%%%%%%%%%%%%%%%%%%%%%%%%%%%%%%%%%%%

%%%%%%%%%%%%%%%%%%%%%%%%%%%%%%%%%%%%%%%%%%%%%%%%%%%%%%%%%%%%%%%%%%%%%%

%%%%%%%%%%%%%%%%%%%%%%%%%%%%%%%%%%%%%%%%%%%%%%%%%%%%%%%%%%%%%%%%%%%%%%


\section*{References}
\begin{thebibliography}{99}

\bibitem{risken1989}
H.~Risken, {\it The Fokker-Planck Equation} (Springer, Berlin, 1989).

\bibitem{aharony1994}
A.~Aharony and D.~Stauffer,
{\it Introduction to Percolation Theory} (Taylor \& Francis, London, 1994).

\bibitem{imry1975}
Y.~Imry and S.~Ma, Phys. Rev. Lett. {\bf 35}, 1399 (1975).

\bibitem{imbrie1984}
J.Z.~Imbrie, Phys. Rev. Lett. {\bf 53}, 1747 (1984).

\bibitem{bricmont1987}
J.~Bricmont and A.~Kupiainen, Phys. Rev. Lett. {\bf 59}, 1829 (1987).

\bibitem{aizenman1989}
M.~Aizenman and J.~Wehr, Phys. Rev. Lett. {\bf 62}, 2503 (1989).

\bibitem{aizenman1990}
M.~Aizenman and J.~Wehr, Commun. Math. Phys. {\bf 130}, 489 (1990).

\bibitem{anderson1958}
P.~W.~Anderson, Phys. Rev. {\bf 109}, 1492 (1958).

\bibitem{nagaoka1982}
Y.~Nagaoka and H.~Fukuyama (Eds.), 
{\it Anderson Localization},
Springer Series in Solid State Sciences 39
(Springer, Berlin, 1982). 

\bibitem{ando1988}
T.~Ando and H.~Fukuyama (Eds.), 
{\it Anderson Localization}, 
Springer Proceedings in  Physics 28 (Springer, Berlin, 1988). 

\bibitem{vantiggelen1999}
B.~van~Tiggelen, in {\it Wave Diffusion in Complex Media},
lecture notes at Les Houches 1998, edited by J.P.~Fouque, NATO
Science (Kluwer, Dordrecht, 1999).

\bibitem{akkermansbook}
E.~Akkermans and G.~Montambaux,
{\it Mesoscopic Physics of Electrons and Photons} (Cambridge University press, 2006).

\bibitem{giamarchi1988}
T.~Giamarchi and H.J.~Schulz,
Phys. Rev. B {\bf 37}, 325 (1988).

\bibitem{fisher1989}
M.P.A.~Fisher, P.B.~Weichman, G.~Grinstein, and D.S.~Fisher,
Phys. Rev. B {\bf 40}, 546 (1989).

\bibitem{scalettar1991}
R.T.~Scalettar, G.G.~Batrouni, and G.T.~Zimanyi,
Phys. Rev. Lett. {\bf 66}, 3144 (1991).

\bibitem{parisi1987}
M.~M\'ezard, G.~Parisi, and M.A.~Virasoro, 
{\it Spin Glass and Beyond} 
(World Scientific, Singapore, 1987).

\bibitem{sachdev1999}
S.~Sachdev,
{\it Quantum Phase Transitions} 
(Cambridge University press, Cambridge, 1999).

\bibitem{ashcroft1976}
N.W.~Ashcroft and N.D.~Mermin, {\it Solid State Physics}
(Saunders College Publishing , New York, 1976).

\bibitem{ioffe1960}
A.F.~Ioffe and A.R.~Regel, Prog. Semicond. {\bf 4}, 237 (1960).

\bibitem{mott1961}
N.F.~Mott and W.D.~Towes, Adv. Phys. {\bf 10}, 107 (1961).

\bibitem{thouless1977} 
D.J.~Thouless, Phys. Rev. Lett. {\bf 39}, 1167 (1977).

\bibitem{gang4}
E.~Abrahams, P.W.~Anderson, D.C.~Licciardello, and T.V.~Ramakrishnan,
Phys. Rev. Lett. {\bf 42}, 673 (1979).

\bibitem{izrailev1999}
F.M.~Izrailev and A.A.~Krokhin,
Phys. Rev. Lett. {\bf 82}, 4062 (1999).

\bibitem{izrailev2005}
F.M.~Izrailev and N.M.~Makarov,
J. Phys. A: Math. Gen. {\bf 38}, 10613 (2005).

\bibitem{lsp2007}
L.~Sanchez-Palencia, D.~Cl\'ement, P.~Lugan, P.~Bouyer, G.V.~Shlyapnikov, and A.~Aspect,
Phys. Rev. Lett. {\bf 98}, 210401 (2007).

\bibitem{cornell2001}
E.A.~Cornell and C.E.~Wieman, Nobel lectures, Rev. Mod. Phys. {\bf 74}, 875 (2002).

\bibitem{ketterle2001}
W.~Ketterle, Nobel lectures, Rev. Mod. Phys. {\bf 74}, 1131 (2002).

\bibitem{dalfovo1999}
F.~Dalfovo, S.~Giorgini, L.P.~Pitaevskii, and S.~Stringari,
Rev. Mod. Phys. {\bf 71}, 463 (1999).

\bibitem{pitaevskii2004}
L.P.~Pitaevskii and S.~Stringari,
{\it Bose-Einstein Condensation} (Oxford University press, 2004).

\bibitem{truscott2001} 
A.G.~Truscott, K.E.~Strecker, W.I.~McAlexander, G.B.~Partridge, and R.G. Hulet, 
Science {\bf 291}, 2570 (2001).
 
\bibitem{schreck2001}
F.~Schreck, L.~Khaykovich, K.L.~Corwin, G.~Ferrari, T.~Bourdel, J.~Cubizolles, and C.~Salomon, 
Phys. Rev. Lett. {\bf 87}, 080403 (2001).

\bibitem{hadzibabic2002}
Z.~Hadzibabic, C.A.~Stan, K.~Dieckmann, S.~Gupta, M.W.~Zwierlein, A.~G\"orlitz, and W.~Ketterle, 
Phys. Rev. Lett. {\bf 88}, 160401 (2002).

\bibitem{roati2002}
G.~Roati, F.~Riboli, G.~Modugno, and M.~Inguscio, 
Phys. Rev. Lett. {\bf 89}, 150403 (2002).

\bibitem{giorgini2007}
S.~Giorgini, L.P.~Pitaevskii, and S.~Stringari,
arXiv:0706.3360.

\bibitem{chu1997}
S.~Chu, Nobel lectures, Rev. Mod. Phys. {\bf 70}, 685 (1998).

\bibitem{cct1997}
C.~Cohen-Tannoudji, Nobel lectures, Rev. Mod. Phys. {\bf 70}, 707 (1998).

\bibitem{phillips1997}
W.D.~Phillips, Nobel lectures, Rev. Mod. Phys. {\bf 70}, 721 (1998).

\bibitem{lattices2000}
G.\ Grynberg and C.\ Robilliard,
Phys. Rep. {\bf 355}, 335 (2000).

\bibitem{burger2001}
S.~Burger, F.~S.~Cataliotti, C.~Fort, F.~Minardi, M.~Inguscio, M.~L.~Chiofalo, and M.~P.~Tosi,
Phys. Rev. Lett. {\bf 86}, 4447 (2001).

\bibitem{kramer2002}
M.~Kr\"amer, L.P.~Pitaevskii, and S.~Stringari, 
Phys. Rev. Lett. {\bf 88}, 180404 (2002).

\bibitem{fertig2005}
C.D.~Fertig, K.M.~O'Hara, J.H.~Huckans, S.L.~Rolston, W.D.~Phillips, and J.V.~Porto,
Phys. Rev. Lett. {\bf 94}, 120403 (2005).

\bibitem{trombettoni2001}
A.~Trombettoni and A.~Smerzi, 
Phys. Rev. Lett. {\bf 86}, 2353 (2001).

\bibitem{anker2005}
Th.~Anker, M.~Albiez, R.~Gati, S.~Hunsmann, B.~Eiermann, A.~Trombettoni, and M.K.~Oberthaler,
Phys. Rev. Lett. {\bf 94}, 020403 (2005).

\bibitem{lye2005}
J.E.~Lye, L.~Fallani, M.~Modugno, D.~Wiersma, C.~Fort, and M.~Inguscio, 
Phys. Rev. Lett. {\bf 95}, 070401 (2005).

\bibitem{clement2005}
D.~Cl\'{e}ment, A.F.~Var\'{o}n, M.~Hugbart, J.A.~Retter, P.~Bouyer, L.~Sanchez-Palencia, D.~Gangardt, G.V.~Shlyapnikov, and A.~Aspect, 
Phys. Rev. Lett. {\bf 95}, 170409 (2005).

\bibitem{fort2005}
C.~Fort, L.~Fallani, V.~Guarrera, J.~Lye, M.~Modugno, D.S.~Wiersma, and M.~Inguscio, 
Phys. Rev. Lett. {\bf 95}, 170410 (2005).

\bibitem{schulte2005}
T.~Schulte, S.~Drenkelforth, J.~Kruse, W.~Ertmer, J.~Arlt, K.~Sacha, J.~Zakrzewski, and M.~Lewenstein, Phys. Rev. Lett. {\bf 95}, 170411 (2005).

\bibitem{clement2006}
D.~Cl\'{e}ment, A.F.~Var\'{o}n, J.A.~Retter, L.~Sanchez-Palencia, A.~Aspect, and P.~Bouyer, 
New J. Phys. {\bf 8}, 165 (2006).

\bibitem{horak1998}
P.\ Horak, J.-Y. Courtois, and G. Grynberg, Phys. Rev. A {\bf 58}, 3953 (1998).

\bibitem{grynberg2000}
G.~Grynberg, P.~Horak, and C.~Mennerat-Robilliard,
Europhys. Lett. {\bf 49}, 424 (2000).

\bibitem{leanhardt2003}
A.E.~Leanhardt, Y.~Shin, A.P.~Chikkatur, D.~Kielpinski, W.~Ketterle, and D.E.~Pritchard, 
Phys. Rev. Lett. {\bf 90}, 100404 (2003).

\bibitem{jones2004}
M.P.A.~Jones, C.J.~Vale, D.~Sahagun, B.V.~Hall, C.C.~Eberlein, B.E.~Sauer, K.~Furusawa, 
D.~Richardson, and E.A.~Hinds, J. Phys. B 37, L{\bf 15} (2004).

\bibitem{kraft2002}
S.~Kraft, A.~G\"unther, H.~Ott, D.~Wharam, C.~Zimmermann, and J.~Fort\'agh, 
J. Phys. B {\bf 35}, L469 (2002).

\bibitem{wang2004}
D.~Wang, M.~Lukin, and E.~Demler, Phys. Rev. Lett. {\bf 92}, 076802 (2004).

\bibitem{esteve2004}
J.~Est\`eve, C.~Aussibal, T.~Schumm, C.~Figl, D.~Mailly, I.~Bouchoule, C.I.~Westbrook, 
and A.~Aspect, Phys. Rev. A {\bf 70}, 043629 (2004).

\bibitem{gavish2005}
U.~Gavish and Y.~Castin,
Phys. Rev. Lett. {\bf 95}, 020401 (2005).

\bibitem{paredes2005}
B.~Paredes, F.~Verstraete, and J.I.~Cirac,
Phys. Rev. Lett. {\bf 95}, 140501 (2005).

\bibitem{courteille2006}
Ph.W.~Courteille, B.~Deh, J.~Fort\'agh, A.~G\"unther, S.~Kraft, C.~Marzok, S.~Slama, and C.~Zimmermann, 
J. Phys. B: At. Mol. Opt. Phys. {\bf 39}, 1055 (2006).

\bibitem{goodman1975}
J.W.~Goodman, 
{\it Statistical Properties of Laser Speckle Patterns} in {\it Laser Speckle and Related Phenomena},
J.-C. Dainty ed. (Springer-Verlag, Berlin, 1975).

\bibitem{goodman2007}
J.W.~Goodman,
{\it Speckle Phenomena in Optics: Theory and Applications}
(Roberts \& Company Publishers, Englewood, Colorado, 2007).

\bibitem{jaksch1998}
D.~Jaksch, C.~Bruder, J.I.~Cirac, C.W.~Gardiner, and P.~Zoller,
Phys. Rev. Lett. {\bf 81}, 3108 (1998).

\bibitem{jaksch2005}
D. Jaksch and P. Zoller,
Ann. Phys. {\bf 315}, 52 (2005).

\bibitem{greiner2002}
M.~Greiner, O.~Mandel, T.~Esslinger, T.W.~H\"ansch, and I.~Bloch,
Nature (London) {\bf 415}, 39 (2002).

\bibitem{lewenstein2007} 
M.~Lewenstein, A.~Sanpera, V.~Ahufinger, B.~Damski, A.~Sen(de), and U.~Sen,
Avd. Phys. {\bf 56}, 243 (2007).

\bibitem{roth2003} 
R. Roth and K. Burnett, Phys. Rev. A {\bf 68}, 023604 (2003).

\bibitem{damski2003}
B.~Damski, J.~Zakrzewski, L.~Santos, P.~Zoller, and M.~Lewenstein,
Phys. Rev. Lett. {\bf 91}, 080403 (2003).

\bibitem{fallani2007}
L.~Fallani, J.E.~Lye, V.~Guarrera, C.~Fort, and M.~Inguscio,
Phys. Rev. Lett. {\bf 98}, 130404 (2007).

\bibitem{sanpera2004}
A.~Sanpera, A.~Kantian, L.~Sanchez-Palencia, J.~Zakrzewski, and M.~Lewenstein,
Phys. Rev. Lett. {\bf 93}, 040401 (2004).

\bibitem{ahufinger2005}
V.~Ahufinger, L.~Sanchez-Palencia, A.~Kantian, A.~Sanpera, and M.~Lewenstein,
Phys. Rev. A {\bf 72}, 063616 (2005).

\bibitem{sanpera2004bis}
L.~Sanchez-Palencia, V.~Ahufinger, A.~Kantian, J.~Zakrzewski, A.~Sanpera, and M.~Lewenstein, 
J. Phys. B: At. Mol. Opt. Phys. {\bf 39}, S121 (2006).

\bibitem{lenoble2004}
O.~Lenoble, L.A.~Pastur, and V.A.~Zagrebnov,
C. R. Physique {\bf 5}, 129 (2004).

\bibitem{falco2007}
G.M.~Falco, A.~Pelster, and R.~Graham,
Phys. Rev. A {\bf 75}, 063619 (2007).

\bibitem{martino2005}
A.~De~Martino, M.~Thorwart, R.~Egger, and R.~Graham,
Phys. Rev. Lett. {\bf 94}, 060402 (2005).

\bibitem{lsp2006}
L.~Sanchez-Palencia, Phys. Rev. A {\bf 74}, 053625 (2006).

\bibitem{lugan2007a}
P.~Lugan, D.~Cl\'ement, P.~Bouyer, A.~Aspect, M.~Lewenstein, and L.~Sanchez-Palencia,
Phys. Rev. Lett. {\bf 98}, 170403 (2007).

\bibitem{yukalov2007}
V.I.~Yukalov, E.P.~Yukalova, K.V.~Krutitsky, and R.~Graham,
Phys. Rev. A {\bf 76}, 053623 (2007).

\bibitem{bilas2006}
N.~Bilas and N.~Pavloff, Eur. Phys. J. D {\bf 40}, 387 (2006).

\bibitem{lugan2007b}
P.~Lugan, D.~Cl\'ement, P.~Bouyer, A.~Aspect, and L.~Sanchez-Palencia,
Phys. Rev. Lett. {\bf 99}, 180402 (2007).

\bibitem{clement2007}
D.~Cl\'ement, P.~Bouyer, A.~Aspect, and L.~Sanchez-Palencia, arXiv:0710.1984 (2007).

\bibitem{chen2007}
Y.P.~Chen, J.~Hitchcock, D.~Dries, M.~Junker, C.~Welford, R.G.~Hulet, arXiv:0710.5187 (2007).

\bibitem{wehr2006}
J.~Wehr, A.~Niederberger, L.~Sanchez-Palencia, and M.~Lewenstein,
Phys. Rev. B {\bf 74}, 224448 (2006).

\bibitem{niederberger2007}
A.~Niederberger, T.~Schulte, J.~Wehr, M.~Lewenstein, L.~Sanchez-Palencia, and K.~Sacha,
Phys. Rev. Lett. {\bf 100}, 030403 (2008).

\bibitem{kuhn2005}
R.~C. Kuhn, C.~Miniatura, D.~Delande, O.~Sigwarth, and C.~A. M{\" u}ller,
Phys. Rev. Lett. {\bf 95}, 250403 (2005).

\bibitem{kuhn2007}
R.~C. Kuhn, C.~Miniatura, D.~Delande, O.~Sigwarth, and C.~A. M{\" u}ller,
New J. Phys. {\bf 9}, 161 (2007).

\bibitem{lsp2005}
L.~Sanchez-Palencia and L.~Santos, 
Phys. Rev. A {\bf 72}, 053607 (2005). 

\bibitem{paul2005}
T.~Paul, P.~Leboeuf, N.~Pavloff, K.~Richter, and P.~Schlagheck,
Phys. Rev. A {\bf 72}, 063621 (2005). 

\bibitem{paul2007}
T.~Paul, P.~Schlagheck, P.~Leboeuf, and N.~Pavloff,
Phys. Rev. Lett. {\bf 98}, 210602 (2007).

\bibitem{olshanii1998}
M.~Olshanii,
Phys. Rev. Lett. {\bf 81}, 938 (1998).

\bibitem{petrov2000}
D.S.~Petrov, G.V.~Shlyapnikov, and J.T.M.~Walraven,
Phys. Rev. Lett. {\bf 85}, 3745 (2000).

\bibitem{kagan1996}
Yu.~Kagan, E.L.~Surkov, and G.V.~Shlyapnikov, 
Phys. Rev. A {\bf 54}, R1753 (1996).

\bibitem{castin1996}
Y.~Castin and R.~Dum, 
Phys. Rev. Lett. {\bf 77}, 5315 (1996).

\bibitem{shepelyanski1994}
D.L.~Shepelyansky,
Phys. Rev. Lett. {\bf 73}, 2607 (1994).

\bibitem{modugno2005}
M.~Modugno,
Phys. Rev. A {\bf 73}, 013606 (2006).

\bibitem{lifshits1988}
I.M.~Lifshits, S.A.~Gredeskul, and L.A.~Pastur, 
{\it Introduction to the Theory of Disordered Systems}, (Wiley and sons, New York, 1988).

\bibitem{gogolin1976a}
A.A.~Gogolin, V.I.~Mel'nikov, and E.I.~Rahba, 
Sov. Phys. JETP {\bf 42}, 168 (1976).

\bibitem{gogolin1976b}
A.A.~Gogolin, Sov. Phys. JETP {\bf 44}, 1003 (1976).

\bibitem{akkermans2006}
E.~Akkermans, S.~Ghosh, and Z.~Musslimani,
cond-mat/0610579.

\bibitem{shapiro2007}
B.~Shapiro, Phys. Rev. Lett. {\bf 99}, 060602 (2007).

\bibitem{skipetov2008}
S.E.~Skipetrov, A. Minguzzi, B.A. van Tiggelen, B. Shapiro,
{arXiv:0801.3631}.

\bibitem{grimm2000}
R.\ Grimm, M.\ Weidem\"uller, and Yu.\ B.\ Ovchinnikov,
Adv. At. Mol. Opt. Phys. {\bf 42}, 95 (2000);
physics/9902072.

\bibitem{huntley1989}
J.M.~Huntley, Appl. Opt. {\bf 28}, 4316 (1989).

\bibitem{massignan2006}
P.~Massignan and Y.~Castin,
Phys. Rev. A {\bf 74}, 013616 (2006).

\bibitem{fallani2005}
L.~Fallani, C.~Fort, J.~Lye, M.~Inguscio,
Opt. Express {\bf 13}, 4303 (2005).

\end{thebibliography}
\end{document}